\documentclass[a4paper,12pt]{article}
\usepackage[cmex10]{amsmath}
\usepackage{latexsym,amssymb}
\usepackage[mathscr]{eucal}
\usepackage{amsxtra,amscd,amsthm,upref,layout,color}
\usepackage{calc}
\usepackage[final]{graphicx}
\def\refup#1{{$^{#1}$}}
\def\p{^{^{\prime}}}\def\pp{^{^{\prime\prime}}}

\def\br{{\bf r}}\def\bbr{\bar{\br}}
\def\hbb{\hat{\bf b}}\def\hbn{\hat{\bf n}}
\def\bE{{\bf E}}\def\bF{{\bf F}}\def\bK{{\bf K}}\def\bR{{\bf R}}

\def\vxip{{\vec{\xi}}_{\perp}}\def\vxipu{{\vec{\xi}}_{1,\perp}}
\def\vxipt{{\vec{\xi}}_{2,\perp}}
\def\rd{{\rm d}}
\def\bRc{{{\bar R}_c}}\def\bRs{{\bar R}_{s}}

\def\oo{{\leavevmode\setbox0=\hbox{h}\dimen0=\ht0 \advance\dimen0
by-1ex\rlap{\raise0.47\dimen0\hbox{\char'27}}o}}
\def\begeq{\begin{equation}}
\def\endeq{\end{equation}}
\def\begdis{\begin{displaymath}}
\def\enddis{\end{displaymath}}

\def\cA{{\cal A}}\def\cC{{\cal C}}
\def\cG{{\cal G}}  
\def\cQ{{\cal Q}}  
\def\cP{{\cal P}}\def\cL{{\cal L}}  
\def\cS{{\cal S}}\def\cV{{\cal V}}

\def\Li{{\rm Li}}
\def\usig{{\underline\sigma}}

\def\GA{{\mathfrak A}}\def\GH{{\mathfrak H}}\def\GK{{\mathfrak K}}\def\GP{{\mathfrak P}}\def\GS{{\mathfrak S}}
\def\arctan{\rm{arctan}}\def\arcos{\rm{arcos}}\def\arcsin{\rm{arcsin}}

\def\ie{{\em i.e.}}\def\eg{{\em e.g.}}%\def\ref#1{$^{#1}$}
\def\etal{{\em et al.}}
\def\hw{{\hat \omega}} \def\hnu{{\hat {\nu}}}
  
 \def\hbn{{\hat{\bf n}}}

\begin{document}
%%%   insert here title, author, institution, abstract etc. 
\title{Small-angle scattering behavior of thread-like and film-like  systems}
\author{  %1
{{Salvino Ciccariello$^a$, Pietro Riello$^b$ and Alvise Benedetti$^b$}}\\%Salvino Ciccariello and Pietro Riello}}\\
%%%%%%%%
  \begin{minipage}[t]{0.9\textwidth}
   \begin{flushleft}
\setlength{\baselineskip}{12pt}
{\slshape  {\footnotesize{$^a$Universit\`{a} di Padova,
Dipartimento di Fisica {\em G. Galilei}, Via Marzolo 8, I-35131 Padova, Italy, and 
$^b$Universit\`a  C\`a Foscari Venezia, Department of Molecular Sciences and Nanosystems, 
 Via Torino 155/B, I-30172 Venezia, Italy.
}}}\\
 \footnotesize{salvino.ciccariello@unipd.it}
\end{flushleft}
\end{minipage}
\\[10mm]
}      %1
%\\[10mm]
 \date{\today}
     % Use \shortauthor to indicate an abbreviated author list for use in
     % running heads (you will need to uncomment it).
\maketitle                        % DO NOT DELETE THIS LINE
%\begin{synopsis}
 %................
%\end{synopsis}
 \begin{abstract} \noindent 
{Film-like and thread-like systems are respectively defined by the property that one of the 
constituting homogenous phases has a constant thickness ($\delta$) or 
a constant normal section (of largest chord $\delta$).  The stick probability 
function of this phase, in the limit $\delta\to 0$,  naturally leads to the 
definition of the correlation function (CF) of a surface or of a curve. This CF  
fairly approximates the generating stick probability function in the 
range of distances larger than $\delta$. The surface and the curve  
CFs respectively behave  as $1/r$ and as $1/r^2$ as $r$ 
approaches to zero. This result implies that the 
relevant small-angle scattering intensities 
behave as $\cP_{\cS}/q^2$ or as $\cP_{\cC}/q$ in an intermediate range 
of the scattering vector ($q$) and as $\cP/q^4$ in the outermost 
$q$-range.  Similarly to $\cP$, pre-factors $\cP_{\cS}$ and  $\cP_{\cC}$  
simply depend on some structural parameters. Depending on the scale 
resolution  it may happen that a given sample looks thread-like at large scale, 
film-like at  small scale and particulate at a finer one. An explicit example
 is reported.  To practically illustrate the above results, the surface and the curve CFs 
of some simple geometrical shapes have been explicitly evaluated. In particular, 
the CF of the  right circular cylinder is explicitly evaluated.  Its limits, as the height or the 
diameter the cylinder approaches zero, are shown to coincide with the CFs 
of a circle and of a linear segment, respectively. 
 \\    \\   
Synopsis:{\em  The scattering intensities of thread-like  and film-like 
systems respectively  behave as $\cP_{\cC}/q$ and  $\cP_{\cS}/q^2$ in an 
intermediate range of the scattering vector $q$. The $\cP_{\cC}$ and 
 $\cP_{\cS}$  expressions are reported. }   \\  
 Keywords: {film-like systems, thread-like systems, surface 
correlation function, curve correlation functions, scale resolution, small-angle 
scattering intensity behavior} }
\end{abstract}
\vfill
%\rightline{DFPD 2015/CM/9}
\eject
{}{}
%.............................................................................80
%%% \end{document}
\def\kb{{k_{_{B}}}}
\def\Na{{N_A}}\def\Nb{{N_B}}\def\mua{{\mu_A}}\def\mub{{\mu_B}}\def\muf{{\mu_F}}
\def\Vf{{V_F}}\def\Tf{{T_F}}
\def\za{{z_A}}\def\zb{{z_B}}\def\zf{{z_F}}
\def\Ta{{T_A}}\def\Tb{{T_B}}\def\Sa{{S_A}}\def\Sb{{S_B}}
\def\Va{{V_A}}\def\Vb{{V_B}}\def\Vig{V{_{ig}}}\def\Tig{T_{ig}}\def\zig{z_{ig}}\def\pa{{p_A}}
\def\pb{{p_B}}\def\Ua{{U_A}}\def\Fa{{F_A}}\def\Ga{{G_A}}\def\Ha{{H_A}}\def\gpa{{\Omega_A}}
\def\Ub{{U_B}}\def\Fb{{F_B}}\def\Gb{{G_B}}\def\Hb{{H_B}}\def\gpb{{\Omega_B}}
\def\ga{{g_{_A}}}\def\gb{{g_{_B}}}\def\siga{{\sigma_{_A}}}\def\sigb{{\sigma_{_B}}}
\def\ma{{m_{_A}}}\def\mb{{m_{_{B}}}}
\def\muhr{{\mu_{_{hr}}}}\def\nhr{{n_{_{hr}}}}\def\muig{{\mu_{_{ig}}}}\def\nig{{n_{_{ig}}}}
\def\pig{{p_{_{ig}}}}
\def\gphr{{\Omega_{_{hr}}}}\def\gpig{{\Omega_{_{ig}}}}\def\gpigD{{\Omega_{_{ig,\,D}}}}
\def\gpigt{{\Omega_{_{ig,\,3}}}}\def\gpigu{{\Omega_{_{ig,\,1}}}}
\def\lamhr{{\lambda_{_{hr}}}}\def\lamig{{\lambda_{_{ig}}}}\def\lam{{\lambda}}
\def\Nhr{N_{_{hr}}}\def\Nig{N_{_{ig}}}\def\Shr{S_{_{hr}}}\def\Sig{S_{_{ig}}}
\def\Lhr{L_{_{hr}}}\def\Lig{L_{_{ig}}}\def\Thr{T_{_{hr}}}\def\Tig{T_{_{ig}}}
\def\phr{p_{_{hr}}}\def\pig{p_{_{ig}}}\def\sigab{\sigma{_{AB}}}\def\sigba{{\sigma_{_{BA}}}}
\def\Li{{\rm Li}}
\def\gb{{g_{_{B}}}}\def\gf{{g_{_{F}}}}\def\te{{\it  {te}}}
%%%%%%%%%%%         INTRODUCTION      $ 1
\section{Introduction}  
Materials characterized by one film-like  or by one thread-like phase are since 
long known. Examples of the first kind are vesicles in solutions or 
oil-water-surfactant systems. Examples of the second kind are polymers 
or amyloid proto-filaments (Avdeev \etal, 2013) in solution.  Small-angle 
scattering (SAS) is one the most suitable tools to characterize these materials 
when the film thickness or the thread diameter is of the nanometer order. 
In fact, based on the knowledge that the SAS intensity [$I(q)$] of a circular 
cylinder decreases, at intermediate scattering vector ($q$) values, as $1/q^2$  
or as $1/q$  depending on whether the cylinder's radius is much larger 
or much smaller then its height  (Porod, 1982; Kirste \& Oberth\"ur, 1982),  
one looks for the existence of a plateau in the plot of  $q^2I(q)$ or of $qI(q)$ 
versus $q$ to conclude that the sample respectively contains ({\em plane}) flat  
particles or ({\em straight}) rod-like ones. 
Actually this conclusion has a more general validity since 
it also applies to the cases of {\em curved}  film-like  and  thread-like phases. To the authors'
knowledge, a first attempt  to get  this more general result,  
starting from the basic equations of 
SAS theory (Guinier \& Fournet, 1955; Kostorz, 1979; Feigin \& Svergun, 1987),   
was done by Teubner (1990)   in the film-like case. \\ 
To make this derivation more detailed as well as to extend it to the thread-like 
case is the aim of this paper according to the following plan.  
Section 2 reports the basic definitions and results of SAS theory for 
three phase samples. Section 3 considers the case where one of the 
constituting phases is a film of thickness $\delta$. One shows how the limit 
$\delta\to 0$ of the stick probability function (SPF) of the film-like phase 
leads to the definition of a {\em surface  correlation function} 
[$\gamma_{\cS}(r)$].  This $\gamma_{\cS}(r)$ function fairly approximates the generating SPF 
in the range $r>\delta$ and it  behaves as $1/r$ at small distances. 
Furthermore, the surface CF  is explicitly evaluated in the case of a sphere 
(\S\,3.4.1), a circle (\S\,3.4.2), a rectangle (\S\,3.4.3) and a cubic surface 
(appendix B).  The circle is the limit of a right circular cylinder as the 
cylinder's height goes to zero. Hence, the corresponding limit of the cylinder 
CF must coincide with that of the circle. To verify this point one needs to 
know the explicit expression of  the cylinder  CF while the only  chord-length 
distribution is presently known (Gille, 2014).  This CF calculation is  
performed in closed form in terms of two  elliptic integral functions (see   
 appendix A).  Section 4 analyzes the case of a tread-like phase. 
In the limit of vanishing thread diameter, the corresponding limit of the 
relevant SPF leads to the definition of the {\em curve correlation function} 
[$\gamma_{\cC}(r)$] that  behaves as $1/r^2$ as $r\to 0$ and fairly 
approximates the considered SPF if  $r>\delta$. The curve CFs of a linear 
segment and a circle are explicitly worked out in \S\,4.1.1 and \S\,4.1.2. 
In \S\,4.1.1 one also shows that the linear segment CF coincides with the 
limit of the cylinder CF as the cylinder's radius goes to zero. 
The behaviors of the scattering intensities relevant to  film-like and thread-like 
phases are discussed in \S\,5. The existence of  an intermediate $q$-range 
where the two intensities respectively behave as $\cP_{\cS}/q^2$ 
and as $\cP_{\cC}/q$  is proved in \S\,5.1 and \S\,5.2. Of course, both intensities 
behave as $\cP/q^4$ in the outer $q$-range. The analytic expressions of  
pre-factors $\cP_{\cS}$ and $\cP_{\cC}$ are also worked out and three simple 
illustrations are reported.  In \S\,5.3 one discusses 
the behavior of the scattering intensity of a right parallelepiped. If the sizes 
of the sides differ more than two order of magnitudes, one finds that the 
intensity at first behaves as  $1/q$, then as $1/q^2$ and finally as $1/q^4$ in agreement 
with the fact that the parallelepiped looks as a thread when observed on a large scale, 
as  a film when the observation scale becomes smaller and, finally,  as a parallelepiped when 
the observation scale has become fine enough to resolve all the involved lengths.  
Section 6 draws the final conclusions. 
%%%%%%%%%%%%%%%%%%%%%%%%%% Basic definitions and properties of the bulk SPFs
\section{Basic definitions and properties of the  SPFs}
Consider a statically isotropic sample made up of three homogeneous phases. The $i$th 
of these occupies  the spatial set $\cV_i$, of volume $V_i$, and has scattering  density 
$n_i$. The total sample occupies the union set $\cV=\cV_i\cup\cV_2\cup\cV_3$ of 
volume $V$.  The volume fraction of the $i$th phase is $\phi_i\equiv V_i/V$. 
The correlation function of the sample is given by (Ciccariello and Riello, 2007) 
\begeq\label{2.5}
\gamma(r)=\sum_{i=1}^3\frac{n(i;j,k)\phi_i(1-\phi_i)}
{2\langle\eta^2\rangle}\Gamma_i(r),\quad i\ne j\ne k. 
\endeq
Here $n(i;j,k)$ denotes  the scattering contrast between phase $i$ and the remaining 
pair of phases $(j,k)$. It is  defined as 
\begeq \label{2.1}
n(i;j,k)\equiv (n_i-n_j)^2+(n_i-n_k)^2-(n_j-n_k)^2. 
\endeq  
$\langle\eta^2\rangle$ is the mean square 
scattering density fluctuation. It is equal to $\sum_{1\le i<j\le3}(n_i-n_j)^2\phi_i\phi_j$ or 
to $\sum_{i=1}^3n(i;j,k)\phi_i(1-\phi_i)/2$. Finally, according to the definition  
\begeq\label{2.4}
\Gamma_i(r)\equiv\frac{P_{i,i}(r)-{\phi_i}^2}{\phi_i(1-\phi_i)}, \ i=1,2,3,
\endeq 
$\Gamma_i(r)$ denotes the CF of the $i$th phase. It is  
%Quantity $\Gamma_i(r)$ is 
determined by the only geometry of the $i$th phase  because $P_{i,i}(r)$ 
is the  {\em stick probability function} (SPF) relevant to the phase pair $(i,i)$. The SPFs were first 
introduced by  Debye \etal\, (1957) [see, also, Peterlin (1965) and Goodisman 
\& Brumberger (1971)], in analogy with the Patterson function, according to the  
definition
\begeq\label{2.3}
P _{i,j}(r)\equiv \frac{1}{4\pi V}\int \rd\hw \int_{R^3} \rho_i(\br_1)
\rho_j(\br_1+r\hw)\rd v_1,\quad i,j=1,2,3.
\endeq
%\begeq\label{2.2}
%\langle\eta^2\rangle = \sum_{1\le i<j\le3}(n_i-n_j)^2\phi_i\phi_j=
%\frac{1}{2}\sum_{i=1}^3n(i;j,k)\phi_i(1-\phi_i). 
%\endeq 
Here $\rho_i(\br)$ is the characteristic function of the set $\cV_i$ (i.e.\ 
$\rho_i(\br)$ is equal to one or zero depending on whether the tip of 
$\br$ falls inside or outside $\cV_i$) and $\hw$ denotes a unit vector 
that spans all possible directions.  The first integral, it being  
an angular average,  ensures  the assumed isotropy of the sample. 
The probabilistic meaning of the SPFs defined by (\ref{2.3}) implies  
(Goodisman \& Brumberger,  1971) that 
\begeq\label{2.3a}
P_{i,j}(\infty)=\phi_i\,\phi_j \quad{\rm and}\quad P_{i,j}(0)=\phi_i\delta_{i,j},
\endeq 
where $\delta_{i,j}$ is Kronecker's symbol. Once the above equalities are substituted within
equations (\ref{2.5}) and (\ref{2.4}) one finds that, whatever $i$, 
\begeq\label{2.7}
\gamma(0)=\Gamma_i(0)=1\quad{\rm and}\quad \gamma(\infty)=\Gamma_i(\infty)=0.
\endeq 
Definitions (\ref{2.5}) and (\ref{2.4}) imply  that the $r$-dependence of the sample CF is determined 
by that of the SPFs that will now briefly reviewed. To this aim it
is observed that  the SPFs  can also be written as (Ciccariello \etal\, 1981)
\begeq\label{2.3b}
P _{i,j}(r) = \frac{1}{4\pi V}\int \rd\hw \int_{\cV_i} \rd v_1\int_{\cV_j} \rd v_2
\delta_3(\br_1+r\hw-\br_2),
\endeq
where $\delta_3(\cdot)$, as specified by the index value,  denotes the 
three-dimensional Dirac function.  From this expression follows that 
the behavior of $P_{i,i}(r)$, as $r\to 0$, mainly reflects the geometrical 
features of surface $\cS_{i}$ that separates the  $i$th  phase from 
the remaining ones.  In fact, it results that 
\begeq\label{2.3c}
P _{i,i}(r) \approx {\GP}_{i}(r)\equiv\phi_i - \frac{r\,S_{i}}{4V} +
\frac{r^2\, {\GA}_{i}}{12\pi\,V}
+\frac{r^3}{24\,V}\Bigl[{\GK}_i+{\mathfrak S}_i\Bigr] ,
%\quad {\rm as}\quad r\approx 0, 
\endeq     %\end{document}
where $S_i$ denotes the area of  surface $\cS_i$ (Porod, 1951), $L_i$ is the total 
length of edges $\cL_i$ present on $\cS_i$ and, finally,  
${\GA}_i$,  $\GK_i$ and ${\mathfrak S}_i$ respectively are the {\em angularity} 
(M\'ering \& Tchoubar, 1968; Porod, 1967; Ciccariello, 1984), the  {\em curvosity} 
(Kirste \& Porod, 1962) and the {\em sharpness} (Ciccariello \& Sobry,  1995)
 of the $\cS_i$ surface. These quantities are defined as follows 
\begeq\label{2.3d}
{\GA}_i =\int_{\cL_i}
\Bigl[1+(\pi-\alpha_i(\ell))\cot\bigl(\alpha_i(\ell)\bigr)\Bigr]\rd\ell +
\sum_{J}\frac{4\pi^2}{{{\GH}_J}^{1/2}}, 
\endeq   %\end{document}
\begeq\label{2.3e}
{\GK}_i = \frac{1}{2}
\int_{\cS_i}\Bigl[3\,H^2(\br)-K_G(\br)\Bigr]\rd S 
\endeq
and
\begeq\label{2.3f}
{\GS}_i =\sum_{I}{\sum_{\imath,\jmath,\ell}}'(-1)_{\imath,\jmath,\ell}
{\mathscr V}_{\imath;\jmath,\ell}(\gamma_{\imath,\jmath},\gamma_{\imath,\ell},\alpha_{\imath}),
\endeq
with 
\begin{eqnarray}\nonumber
&&
{\mathscr V}_{\imath;\jmath,\ell}(\gamma_{\imath,\jmath},\gamma_{\imath,\ell},\alpha_{\imath})=
-\frac{1}{4}\Bigl(1+2\pi\,(\cot\alpha_{\imath}/\sin\alpha_{\imath})(\cot\gamma_{\imath,\jmath}+
\cot\gamma_{\imath,\ell})+\\
&&\quad\quad (\pi-\gamma_{\imath,\jmath})\bigl[\cot\gamma_{\imath,\jmath}-
2(\cot\alpha_{\imath}\cot\alpha_{\jmath})/\sin\gamma_{\imath,\jmath})\bigr]+\label{2.3g}\\
&&\quad\quad (\pi-\gamma_{\imath,\ell})\bigl[\cot\gamma_{\imath,\ell}-
2(\cot\alpha_{\imath}\cot\alpha_{\ell})/\sin\gamma_{\imath,\ell})\bigr]-\nonumber\\
&&\quad\quad(\pi-\gamma_{\jmath,\ell})\bigl[\cot\gamma_{\jmath,\ell}-
2(\cot\alpha_{\imath}\sin\gamma_{\jmath,\ell})/(\sin\gamma_{\imath,\jmath}
\sin\gamma_{\imath,\ell}
\sin\alpha_{\imath})\bigr]\Bigr).\nonumber
\end{eqnarray}  
In (\ref{2.3d}), the first contribution (Ciccariello \etal, 1981) refers to the edges and 
$\alpha_i(\ell)$ denotes 
the dihedral angle value at the edge point with curvilinear coordinate $\ell$.   
The second contribution (Ciccariello \& Benedetti, 1982) arises from possible points of 
contact between  different branches 
of surface $\cS_i$. These points are indexed by $J$, and ${\GH}_J$ denotes the 
corresponding value of the Hessian of the function resulting from the difference of 
the two branches.  Kirste \& Porod (1962) obtained  equation (\ref{2.3e}). Here 
$H(\br)$ and $K_G(\br)$ respectively denote the mean and the Gaussian 
curvatures of surface $\cS_i$ at the point (with position vector) $\br$.  
The two curvatures  are related to the minimum [$R_m(\br)$] and the 
maximum  [$R_M(\br)$] curvature radius of the surface by
\begin{eqnarray}
H(\br) &=&\frac{1}{2}\Bigl(\frac{1}{R_m(\br)}+\frac{1}{R_M(\br)}\Bigr),\label{2.12}\\
K_G(\br) &=&\frac{1}{R_m(\br)\,R_M(\br)}.\label{2.13}
\end{eqnarray}
Finally, equation (\ref{2.3f}) was obtained by Ciccariello \& Sobry (1995). The outer 
summation there present is performed over all the vertex points of $\cS_i$ and the inner one over 
all distinct triple of edges that enter into the $I$th vertex. Further, 
$\gamma_{\imath,\ell}$ denotes the angle formed by the $\imath$th and the $\ell$th 
edge while $\alpha_{\imath}$ 
is the dihedral angle at the $\imath$th edge. It is also noted that, whenever a 
couple of facets do not share a common edge but only a vertex, one has to 
consider the virtual edge along which the facets, once they are prolonged, 
intersect each other. This fact explains the presence in (\ref{2.3g}) of the sign 
factor $(-1)_{\imath,\jmath,\ell}$. One  should refer to page 65 of  the above paper 
for the full definition. This paper also shows that the SPF of a 
phase bounded by a polyhedral surface is a 3rd degree polynomial within the 
innermost range of distances, \ie\, it  is exactly given by equation (\ref{2.3c}) 
with $H=K_G=0$.\\ 
A further property of SPFs is the fact that, whenever parts of the 
interfaces are parallel to each other at a relative orthogonal distance $\delta$, 
the second derivative of the relevant SPF shows a finite  (Wu \& Schmidt, 1974) 
or a logarithmic discontinuity (Ciccariello, 1989) as $r\to \delta$. This 
property was thoroughly discussed by  Ciccariello (1991) and is practically useful 
(see, \eg\,, Melnichenko \& Ciccariello, 2012 and references therein).  
This property applies to  film-like samples and yields an oscillatory contribution 
decreasing as $q^{-4}$ in the range $q>2\pi/\delta$.  For simplicity,  however,   this 
contribution will not  explictly be considered in  \S\,5.1.\\ 
%%%%%        \end{document}
%The CF of the full sample reads
%\begeq\label{2.5}
%\gamma(r)=\sum_{i=1}^3\frac{n(i;j,k)\phi_i(1-\phi_i)}{2\langle\eta^2\rangle}\Gamma_i(r). 
%\endeq
%From equation (\ref{2.4}) immediately follows that 
%\begeq\label{2.7}
%\Gamma_i(0)=1 ,\quad {\rm and}\quad \Gamma_i(\infty)=0,\quad i=1,2,3,
%\endeq
%and from equations (\ref{2.5}) and (\ref{2.2}) that 
%\begeq\label{2.8}
%\gamma(0)=1 ,\quad {\rm and}\quad \gamma(\infty)=0. 
%\endeq 
%%%%%%%%%%%%%%%%%%%%%%%%%%%%%%%%%%%%%%%%%%%%%%%%%%%%%%
%%%%%%%\subsection{Film-like phase and surface correlation function}
\section{Film-like phase and surface CF}
%%%%%%%%%%%%%%%%%%%%%%%%%%%%%%%%%%%%%%%%%%%%%%%%%%%%%%%%
Assume now that  phase 1 have  a film-like structure, \ie\   $\cV_1$ is a connected 
or disconnected  set delimited by a left  ($\cS_{1,l}$)  and a right  ($\cS_{1,r}$) surface 
at a relative constant orthogonal distance $\delta$. Moreover, one also assumes 
that $\cS_{1,l}$  and $\cS_{1,r}$ be smooth (\ie\, they have no edges, vertices and contact points) 
and  that $\delta$ be small in the sense that it obeys the inequality 
\begeq\label{3.1}
\delta << \bRs \equiv\langle (1/R_m^2+1/R_M^2)\rangle^{-1/2},
\endeq   
where $\bRs$ denotes the mean  value of the curvature radii  of  
$\cS_{1}=\cS_{1,l}\cup\cS_{1,r}$, the surface bounding $\cV_1$.   Denote  the 
surface midway $\cS_{1,l}$ and $\cS_{1,r}$  by $\cS_{f}$. Then, $\br_1$, the 
position vector of  a generic point P  of $\cV_1$, can uniquely be written as 
\begeq\label{3.2}
\br_1 =\bbr_1+\xi\, \hnu(\bbr_1),
\endeq  
where $\bbr_1$  denotes the position of the point intersection  of $\cS_f$ with 
the straight line [with direction $\hnu(\bbr_1)$] that going through P,  is normal 
to $\cS_f$.  \ $\xi$ is the orthogonal distance  of P from $\cS_f$. In equation 
(\ref{2.3b}) with $i=j=1$, the infinitesimal volume element $\rd v_1$ located at $\br_1$ 
can be written,  up to terms  $o(\delta)$,   as [see, \eg,  equation (4.7) of 
Ciccariello (1991)] 
\begeq\label{3.3}
dv_1=[1+\xi H(\bbr_1)]dS_1\,d\xi, \quad -\delta/2\le \xi\le\delta/2,
\endeq 
where $dS_1$ is the infinitesimal surface element of $\cS_f$ located at  
$\bbr_1$.  Using equations  (\ref{3.2}),  (\ref{3.3}) and (\ref{2.3b}), one concludes 
that the SPF of  film-like phase 1 takes the form       %%%% \end{document}
\begin{eqnarray}\nonumber
&& P _{1,1}(r)\equiv \frac{1}{4\pi V}\int_{-\delta/2}^{\delta/2}\rd\xi_1
\int_{-\delta/2}^{\delta/2} \rd\xi_2\,
\int \rd\hw \int_{\cS_f} \rd S_1\int_{\cS_f}\rd S_2\,\bigl(1+\xi_1\,H(\bbr_1)\bigr)\times  \\
&&\quad\quad\quad \quad\quad\quad\bigl(1+\xi_2\,H(\bbr_2)\bigr)
\delta_3\bigl(\bbr_1+\xi_1\hnu(\bbr_1)+r\hw-\bbr_2-\xi_2\hnu(\bbr_2)\bigr).\label{3.4}
\end{eqnarray} 
%%%%   \subsection{The surface CF definition}
\subsection{The surface CF definition}
\begin{figure}[!h]
{\includegraphics[width=7.truecm]{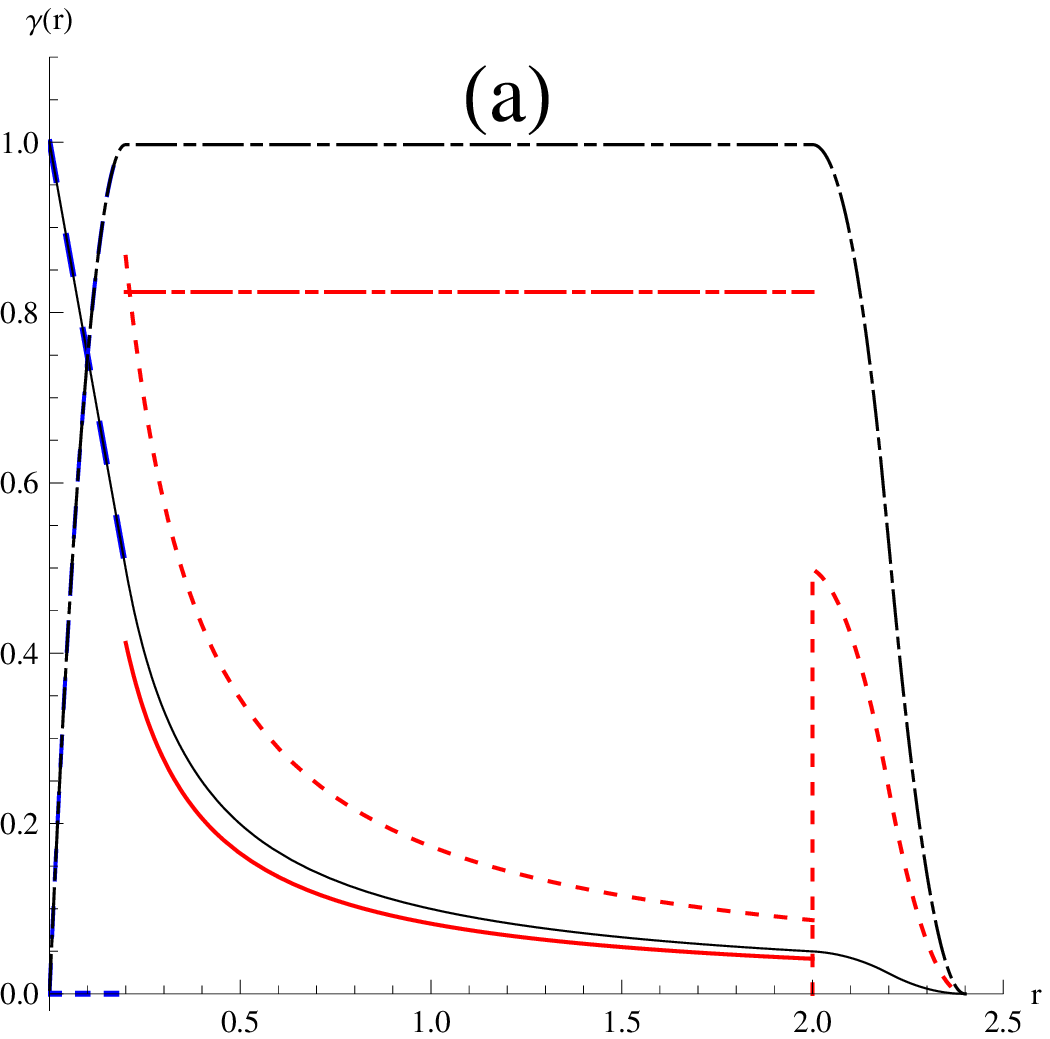}} %%    figure 1a and 1b
{\includegraphics[width=7.truecm]{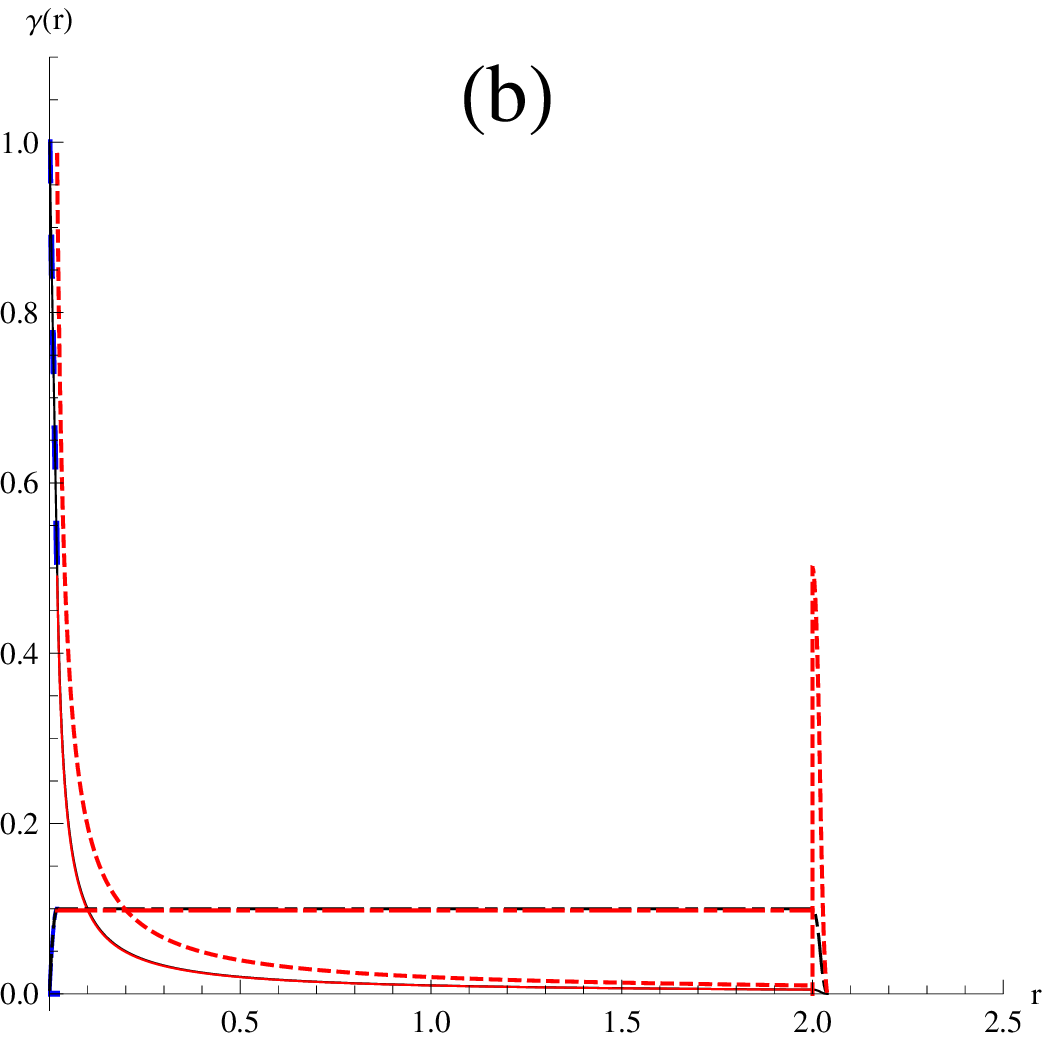}} %%% evaluated in the  file "Spherical_Shell "
\caption{\label{Fig1} {(a): the black continuous line plots  $\gamma_{\cV,ss}(r)$, the CF
of a spherical shell [it is recalled that in the case of a single particle one has  $\phi_1=1$ and 
$P_{1,1}(r)=\gamma_{\cV,ss}(r)$, the last function being given by (\ref{3.25})] 
with inner and outer radii  equal to $R_i=R=1$ and $R_o=1.2$, 
so that $\delta=(R_o-R_i)=0.2$ and  $\delta/{\bar R}=0.25$. [For the 
${\bar R}$ definition see equation (\ref{3.1})]. The continuous red line is the 
plot of  $\delta^2 S_f\gamma_{\cS,sph}(r,R)/V$ within the range $\delta<r<2R$, 
$\gamma_{\cS,sph}(r,R)$ being the CF of a spherical surface of radius $R=1$ 
[see equation (\ref{3.26})]. The thick long-dash  blue  curve is the plot of  
${\mathfrak P}_{1,\cS}(r)$.
The dotted blue (hardly visible) and red curves show the approximation 
discrepancy since they plots  
$c\cdot\bigl(\gamma_{\cV,ss}(r)-{\mathfrak P}_{1,\cS}(r)\bigr)$ and  
$c\cdot\bigl(\gamma_{\cV,ss}(r)-\delta^2 S_f\gamma_{\cS,sph}(r,R)/V\bigr)$ 
with $c=10$. 
The dash-dotted black, blue and red broken lines respectively are the plots 
$10\cdot r\gamma_{\cV,ss}(r)$,  $10\cdot r{\mathfrak P}_{1,s}(r)$ and 
$10\cdot r\,\delta^2 S_f\gamma_{\cS,sph}(r,R)/V$ (red) .  (b): as 
in (a) with $R_i=1$ and $\delta=0.02$ and $c=100$ (the dash-dotted curves merge 
in the single red one). The two panels makes it evident that the approximation 
gets more and more accurate as $\delta/R$ decreases.}}
\end{figure}

In the range of distances $0\,\le\, r\,<\,\delta$ this function is reliably approximated 
by equation (\ref{2.3c}) that in the present case  reads 
\begeq\label{3.5}
P _{1,1}(r) \approx {\mathfrak P}_{1,\cS}(r)\equiv \phi_i - \frac{2\,S_{f}\,r}{4V} +
\frac{r^3\,S_f}{48\,V}\Bigl(\langle 3H^2-K_G\rangle_{\cS_f}\Bigr)\quad  {\rm if}\quad 0<r<\delta,
%\quad 0 \le  r<\delta, 
\endeq 
since the smoothness of the surface and the smallness of $\delta$ makes 
the surface area approximation $S_1\approx 2S_f$ accurate. \\ 
In the range $r>\delta$, the smallness of $\delta$ makes it possible to 
set $\xi_1=\xi_2=0$ within the  integrand of (\ref{3.4}) so that the SPF becomes 
\begeq\label{3.6}
P_{1,1}(r)\approx \frac{\delta^2 S_f}{V}\gamma_{\cS}(r)\quad{\rm if}\quad r>\delta, 
\endeq 
with  
\begin{eqnarray}\label{3.7}
\gamma_{\cS}(r)&\equiv& \frac{1}{4\pi\,S_f }\int \rd\hw\int_{\cS_f}dS_1
\int_{\cS_f}\rd S_2
\delta_3\bigl(\bbr_1+r\hw-\bbr_2\bigr), \  \  r>0.
\end{eqnarray}                %%%\end{document}
Function $\gamma_{\cS}(r)$ is named {\em surface 
correlation function} because it is fully determined by the surface set $\cS_f$. Function 
$\gamma_{\cS}(r)$  was first introduced by Teubner (1990). [Note, however, 
that this author used $\frac{1}{4\pi V}$,  instead of $\frac{1}{4\pi S_f}$,  as normalization 
factor  in front of the integral reported in (\ref{3.7}).] \\ 
Figures \ref{Fig1}  and \ref{Fig2} show  the accuracy achieved in approximating the 
bulk SPF $P_{1,1}{r}$ by the 3rd degree $r$-polynomial (\ref{3.5})  in the range 
$0<r<\delta$ and by the surface CF, via equation (\ref{3.6}),  in the range $r>\delta$ 
for two simple particle shapes: the spherical shell and the cylindrical disk. The 
accuracy drastically improves as $\delta$ gets smaller.  
%%%%   \subsection{Properties of the surface CF}
\subsection{Properties of the surface CF}
From equation (\ref{3.6}) and the fact that $P_{1,1}(r)$ is dimensionless it follows that 
the surface CF has dimension length$^{-1}$. 
The leading behavior of the surface CF as $r\to 0$ was obtained by Teubner (1990),   
under the assumption that the surface is closed [\ie,  with no boundary] and smooth, 
by the same procedure that Kirste \& Porod (1962) followed to get their relation. 
It reads 
\begeq\label{3.8}
\gamma_{\cS}(r)\approx \frac{1}{2 \,r}+
\frac{r}{32}\Big\langle\Big(\frac{1}{R_m}-\frac{1}{R_M}\Big)^2\Big\rangle+o(r),
\endeq
where the angular brackets denote the average over $\cS_f$. 
This relation shows that  the surface CF diverges as $r\to 0$ so that it cannot be 
normalized to one at the origin as it happens for standard CFs. \\
As $r\to\infty$,  the  leading behavior of the surface CF is   
\begeq\label{3.9}
\gamma_{\cS}(r)\approx \frac{{\,S_f}}{V}.
\endeq 
This  immediately follows from equations  ({\ref{2.3a}) and (\ref{3.6}) since 
\begeq\nonumber  
\lim_{r\to\infty}\gamma_{\cS}(r)=\lim_{r\to\infty}\frac{V}{\delta^2\,S_f}\,P_{1,1}(r)\approx 
\frac{V\,\phi_1^2}{\delta^2\,S_f}\approx 
 \frac{V\,\,S_f^2\delta^2}{\delta^2\,S_f\,V^2} = \frac{ S_f}{V}. 
\endeq 
It is also noted that the surface CF defined by equation (\ref{3.7}) generalizes the two 
dimensional (2D) CF defined as (Ciccariello, 2009) 
\begeq\label{3.21}
\gamma_2(r)=\frac{1}{2\pi S}\int \rd \hw\int_{\cS}\rd S_1\int_{\cS}\rd S_1
\delta_{2}(\br_1+r\hw-\br_2), 
\endeq
where $\cS$ is a plane  set and $\hw$ is a unit vector that spans the unit circle. 
To get the relation between the 
two quantities in the case where  the $\cS_f$ set,  present in (\ref{3.7}), is  plane,
one chooses the plane where $\cS_f$ lies as the $z=0$ plane.  Then, in (\ref{3.7})  
one writes 
$$\delta_3(\br_1+r\hw-\br_2)\rd \hw= 
\delta_2(\br_1+r\hw_{||}-\br_2)\delta_1(r\cos\theta)\rd \varphi \rd (cos\theta),$$
[where $\hw_{||}$ is the projection of $\hw$ onto the $z=0$ plane] 
and, after integrating over $\theta$, one gets 
\begeq\label{3.22}
\gamma_{\cS}(r)=\frac{1}{4\pi\,S_f\,r}\int\rd \hw  \int_{\cS_f}\rd S_1\int_{\cS_f}\rd S_1
\delta_{2}(\br_1+r\hw-\br_2), 
\endeq
where the outermost integral is performed on the unit circle. 
Comparing (\ref{3.22}) with (\ref{3.21}) and assuming that $\cS_f=\cS$ one finds 
\begeq\label{3.23}
2\,r\,\gamma_{\cS}(r)= \gamma_2(r).
\endeq 
Ciccariello (2010) showed that, as $r\to 0$,  $\gamma_2(r)\approx (1-r\, L/\pi S)$  
where $L$ denotes the length of the boundary of $\cS$. Thus, relation (\ref{3.23}), 
allows us to generalize equation (\ref{3.8}) to the case where $\cS_f$ is an open  
surface,  and has therefore a boundary of perimeter $L_f$, in the following way   
\begeq\label{3.24}
\gamma_{\cS}(r)\approx \frac{1}{2\,r} -\frac{\,L_f}{2\,\pi\,S_f}+
\frac{r}{32}\Big\langle\Big(\frac{1}{R_m}-\frac{1}{R_M}\Big)^2\Big\rangle+o(r)
\quad {\rm as}\quad r\to 0.
\endeq
Finally it is observed that the moments of the surface CF can be expressed in terms of 
the moments of the surface if $\frac{S_f}{V}\approx 0$. . In fact, putting 
\begeq\label{3.25}
M_{S,m}\equiv \int_0^{\infty}r^{2m+2}\gamma_{\cS}(r)dr,
\endeq
from equation(\ref{3.7}) follows
\begin{eqnarray}\label{3.26}
M_{S,m}&=&\frac{1}{4\pi\,S_f }\int r^{2m}\rd v\int_{\cS_f}dS_1
\int_{\cS_f}\rd S_2 \delta_3\bigl(\bbr_1+\br-\bbr_2\bigr)\\
  &=&\frac{1}{4\pi\,S_f }\int_{\cS_f}dS_1
\int_{\cS_f}\rd S_2  (\br_2-\br_1)^{2m}.\nonumber
\end{eqnarray}
The last expression is equal to the sum of products of different moments of  surface $\cS_f$ [see 
appendix B of Ciccariello (2014)]. 
In particular, if $m=0$, one simply finds
\begeq\label{3.26a}
M_{S,0}=\frac{S_f}{4\pi}.
\endeq
%%% \subsection{Integral expressions of the derivatives of the surface CF}
\subsection{Integral expressions of the derivatives of the surface CF}
The comparison of  ({\ref{3.7})  with equation (3.1) of Ciccariello 
\etal\, (1981) shows that the surface CF and the second derivative of $P_{1,1}(r)$ 
nearly have the same mathematical structure since the only  difference, aside from 
the normalization factors,  lies in  the 
angular  contribution $(\hnu(\bbr_1)\cdot\hw)(\hnu(\bbr_2)\cdot\hw)$ that is only 
present in the $P_{1,1}\pp(r)$ expression. Consequently, many of the elaborations 
worked out for this expression apply, {\em mutatis mutandis}, to the surface CF. 
In particular, let $\Sigma(\br_1,r)$ denote the spherical surface with center at 
$\br_1$ and radius $r$. Then equation (\ref{3.7}) can be written as 
\begeq\label{3.10}
\gamma_{\cS}(r)= \frac{1}{4\pi\,S_f \,r^2}\int_{\cS_f}\rd S_1
\int_{\Sigma(\bbr_1,r)} \rd S\int_{\cS_f}
\delta_3\bigl(\bbr_1+r\hw-\bbr_2\bigr)\rd S_2.
\endeq
Omitting the overbar, denoting by $\cC_{\Sigma}(\br_1,r)$ the intersection curve  
of $\Sigma(\br_1,r)$  with $\cS_f$ (it is formed by the points of  $\cS_1$ that are 
at distance $r$ from point $\br_1$  of $\cS_f$) and proceeding as in the just 
mentioned paper, one finds that $\gamma_{\cS}(r)$ takes the form 
%%\end{document}
\begeq\label{3.11}
\gamma_{\cS}(r)= \frac{1}{4\pi\, S_f\,r^2}\int_{\cS_f} \rd S_1\int_{\cC_{\Sigma}(\bbr_1,r)}
\frac{1}{\sqrt{1-\bigl(\hnu(\br_1+r\hw(\ell))\cdot\hw(\ell)\bigr)^2}}\rd\ell. 
\endeq 
In this expression, $\ell$ is the curvilinear coordinate of  the curve points 
(note that the curvilinear coordinate definition is such that the distance 
between two points,  having   curvilinear coordinates $\ell$ and $\ell+d\ell$, 
is equal to $d\ell$), $\hnu\big(\br_1+r\hw(\ell)\big)$ denotes the unit vector 
orthogonal to (and pointing outwardly to) $\cS_f$ at the 
point $\br_2(\ell)\equiv\br_1+r\hw(\ell)$ and $\hw(\ell)$ specifies  the direction 
of  the vector $(\br_2(\ell)-\br_1)$. \\
If one  puts    
\begeq\label{3.12}
\cos\Theta(\br_1,r,\ell)\equiv  \hnu\bigl(\br_1+r\hw(\ell)\bigr)\cdot\hw(\ell),
\endeq 
and         
\begeq\label{3.14}
p(\br_1,r)\equiv \int_{\cC_{\Sigma}(\br_1,r)}\frac{1}{r^2\,\sin\Theta(\br_1,r,\ell)}d\ell, 
\endeq
equation (\ref{3.11}) becomes 
\begeq\label{3.13}
\gamma_{\cS}(r)= \frac{1}{4\pi\, S_f}\int_{\cS_f} {\rd}S_1\int_{\cC_{\Sigma}(\br_1,r)}
\frac{1}{r^2\,\sin\Theta(\br_1,r,\ell)}{\rd}\ell=
\frac{1}{4\pi\,S_f}\big\langle  p(.,r)\big\rangle_{S_f}.
\endeq 
Thus, $\gamma_{\cS}(r)$ is proportional to $\big\langle  p(.,r)\big\rangle_{S_f}$ that 
denotes the average value of $p(\br_1,r)$ over $\cS_f$.     
The $n$th derivative of $\gamma_{\cS}(r)$ takes the form  
\begeq\label{3.15}
{\gamma_{\cS}}^{(n)}(r)=\frac{1}{4\pi\,S_f }\int \rd\hw\int_{\cS_f}\rd S_1\int_{\cS_f}
\rd S_2 (-\hw\cdot\nabla_2)^n\delta\bigl(\bbr_1+r\hw-\bbr_2\bigr), 
\endeq 
which implies that the reduced integral expressions involve an integration, 
along the same  $\cC_{\Sigma}(\br_1,r)$ curve, of an integrand that changes with
 the derivative order $n$. To get the integrand expression, 
proceeding as in  Ciccariello (1995), one denotes by $\bR(u,v)$ the parametric 
equation of $\cS_1$, and  by $\bR_u(u,v)$ and $\bR_v(u,v)$ the corresponding 
derivatives with respect to $u$ and $v$.  
One puts %%%%%\end{document}
\begeq\label{3.16}
{\vec\cA}^{(0)}\bigl(\br_1,\br_2(\ell),r\bigr)\equiv\frac{{\hat\tau}(\ell)} 
{||\bR\bigl(u(\ell),v(\ell)\bigr)-\br_1||^2\, \sin\Theta(\br_1,r,\ell)},
\endeq
[where $\br_2(\ell)\equiv\bR\bigl(u(\ell),v(\ell)\bigr)$ and   
${{\hat\tau}(\ell)}$ denotes the unit vector tangent to $\cC_{\Sigma}(\br_1,r)$ 
at $\br_2(\ell)$] and 
\begeq\label{3.17}
{\vec\cA}^{(n)}\bigl(\br_1,\br_2(\ell),r\bigr)\equiv\frac
{\bR_u\cdot{\vec\cA}^{(n-1)}_v -\bR_v\cdot{\vec\cA}^{(n-1)}_u } 
{||\bR_u\times \bR_v||\, \sin\Theta(\br_1,r,\ell)}{\hat\tau}(\ell),\quad n=1,2,\ldots
\endeq
where the reported partial derivatives are evaluated at the argument values shown 
on the left hand side. In this way, the integral expression of the $n$th derivative 
of the surface CF becomes  
\begeq\label{3.19}
{\gamma_{\cS}}^{(n)}(r)= \frac{1}{4\pi\,S_f}\int_{\cS_f} \rd S_1\int_{\cC_{\Sigma}(\br_1,r)}
\bigg({\vec\cA}^{(n)}\bigl(\br_1,\br_2(\ell),r\bigr)\cdot {\hat \tau}(\ell)\bigg) 
{\rd}\ell 
\endeq
under the assumption that the involved partial derivatives exist. \\
%%%%%%%%%      \subsection{Examples of surface CFs}     %%%%%%%%%%%%%
%%%%%%%%%%     \subsubsection{The spherical surface}   %%%%%%%%%%%%
\subsection{Examples of surface CFs}
The following subsections report the surface CFs relevant to a spherical surface, 
a circle and  a rectangle, while the surface CF of  a cubic surface is worked out in 
appendix B.
\subsubsection{The CF of spherical surface}
%%%%%%%%%%%%%%%%%%%%%%%%%%%%%%%%%%%%%%%%%%%%%%%%%%%%%%%%%%
Consider a spherical shell of inner and outer radius equal to $R$ and  
$R+\delta$ with $\delta<R$. The bulk CF was calculated  by Glatter (1982) (see also 
Fedorova \& Emelyanov, 1977) and  reads 
\begeq\label{3.25}
\gamma_{\cV,ss}(r)=
\begin{cases}
  \frac{\pi \big(r^3 - 6 r (2 R^2 + 2 R \delta + \delta^2) + 
      8 \delta (3 R^2 + 3 R \delta + \delta^2)\big)}{6V} & \text{if }  0 < r < \delta, \\
\frac{\pi \delta^2 (2 R + \delta)^2}{2\, r\,V}  & \text{if }  \delta < r < 2R, \\
\frac{\pi ( 12 r^2 R^2-r^4  - 16 r R^3 + 
      6 \delta^2 (2 R + \delta)^2)}{12\, r\,V}  & \text{if }  2R < r < 2R+\delta, \\
\frac{\pi (r - 2 (R + \delta))^2 (r + 4 (R + \delta))}{12\,V}
& \text{if }  2R+\delta < r < 2R+2\delta,\\
0 & \text{if }  2R+2\delta < r,
\end{cases}
\endeq
with $V=4\pi \big( (R+\delta)^3-R^3\big)/3$.\\  
As $\delta\to 0$, the spherical shell  becomes a spherical surface of radius R. 
The corresponding CF is easily evaluated by (\ref{3.13}).  In fact, $\cC_{\Sigma}(\br_1,r)$ 
is a circle of radius $R_1=r\,\sqrt{1-r^2/4R^2}$ and $\cos\Theta(\br_1,r,\ell)=r/2R$. 
In this way,  from (\ref{3.13}) follows that 
\begeq\label{3.26}
\gamma_{\cS,sph}(r,R)=
\begin{cases} 
\frac{1}{2\,r} & \text{if}\quad   0 < r < 2R,\\
0 & \text{if}\quad 2R<r.
\end{cases}
\endeq
This expression is equal to $\lim_{\delta\to 0}V\gamma_{\cV,ss}(r)/S_f\delta^2$ as 
required by equation (\ref{3.6}).  
Figures \ref{Fig1}a and \ref{Fig1}b allows one to appreciate the accuracy achievable 
in approximating $\gamma_{\cV,ss}(r)$ by 
$\delta^2\,S_f \gamma_{\cS,sph}(r,R)/V$  for the cases $(R,\,\delta)=(1,\,0.2)$u and 
$(1,\,0.02)$u, respectively.
%%%%%                   \end{document}
%%%%%     The surface CF of a circle     
\subsubsection{The surface CF of a circle}
Consider a circular cylinder of radius $R$ and height $h$. If one lets $h$ go to zero, 
the cylinder shrinks to a circle of radius $R$.  The surface  CF  of the circle is easily 
evaluated by equation (\ref{3.23}).  In fact the value of the 2D CF of a circle at a given 
distance $r$ is proportional to the overlapping area of the outset circle and the  
circle shifted by $r$.  In this way, accounting for the appropriate proportionality 
constant, one finds that the surface CF $\gamma_{\cS,crcl}(r,R)$ of a circle is 
\begin{figure}[!h]
{\includegraphics[width=7.truecm]{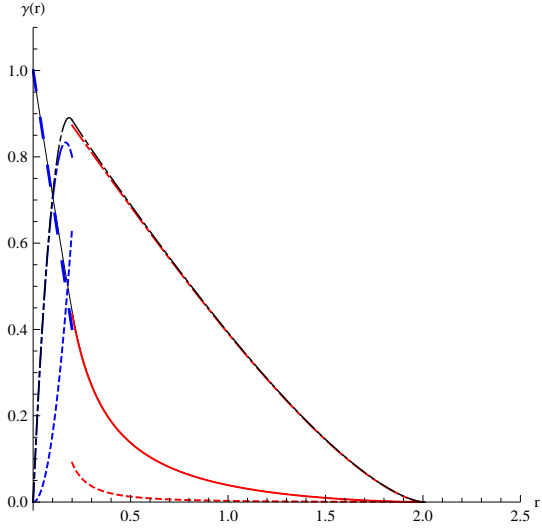}} %%    DISK-LIKE 
%% see MATHEMATICA file 
%  /Users/salvino/Desktop/WORK_IN_PRGS/THREAD_FILM_LIKE _SYSTEMS/cylinder_NEW_a
\caption{\label{Fig2} {The shown curves have the same meaning of those reported 
in  Fig. \ref{Fig1}.  They do however refer to a circular cylinder of radius $R=1$u and height 
$h=0.2$u.  The black, blue  and red dash-dotted  curves are the plots of 
$10\cdot r\,\gamma_{dsk}(r,R,h)$ [see equation(\ref{A.5})],   $10\cdot r\,{\mathfrak P}_{1,\cS}(r)$ 
 [see Eq. (\ref{3.5})] and 
$10\cdot r\,h^2 S_f\gamma_{\cS,crcl}(r,R)/V_{cyl}$ [see Eq. (\ref{3.27})].  The dotted curves plot 
the approximation error multiplied by $c=15$. 
 }} 
\end{figure}   
\begeq\label{3.27} 
\gamma_{\cS,crcl}(r,R)=
\begin{cases}
-\frac{\sqrt{4R^2-r^2}}{4\,\pi\,R^2}+
\frac{1}{\pi\,r}\arccos\frac{r}{2R} &\text{if}\quad  0<r<2R,\\
0  &\text{if}\quad  2R <r.
\end{cases}
\endeq 
One easily verifies that 
\begeq\nonumber
\lim_{h\to 0} \frac{V_{cyl}}{h^2\,S_{f}}\gamma_{dsk}(r, R,h) = \gamma_{\cS,crcl}(r,R),
\endeq 
where $\gamma_{dsk}(r, R,h)$ denotes the cylinder CF that is given by 
 (\ref{A.5}). Besides, the small distance expansion of (\ref{3.27}) 
yields 
\begeq\label{3.27bis} 
\gamma_{\cS,crcl}(r,R)\approx \frac{1}{2 r} -\frac{1}{\pi\,R} + 
\frac{r^2}{24\, \pi\, R^3}+O(r^3),
\endeq 
in agreement with equation (\ref{3.24}).\\ 
Similarly to Fig.\,\ref{Fig1}, Fig.\,\ref{Fig2} shows the accuracy achieved in 
approximating $\gamma_{dsk}(r, R,h)$, the cylinder CF, by the relevant 
${\mathfrak P}_{1,\cS}(r)$ polynomial within the range $0<r<h$  and by  
$h^2 S_f\gamma_{\cS,crcl}(r,R)/V_{cyl}$  if $r>h$ in the case where 
the cylinder has height equal to $h=0.2$u  and radius equal to $R=1$u. 
%%%%%    {The surface CF of a rectangle     %%%%%%%%%%%%
\subsubsection{The surface CF of a rectangle}
Consider a rectangle of sides $a$ and $b$ with $a<b$. Its surface CF 
$\gamma_{\cS,rct}(r,a,b)$ is easily obtained using equation (\ref{3.23}) and 
recalling that the 2D CF of any plane polygon has an algebraic form (Ciccariello, 
2009).  Hence, using the result of this paper, one finds that the surface CF 
of the aforesaid rectangle is 
\begeq\label{5.3.2}
\gamma_{\cS,rct}(r;a,b)=\begin{cases}
\frac{\pi a b  - 2 (a + b) r + r^2}{2 \pi r a b },\quad\quad\quad\quad\quad\quad  
\quad\quad  0 < r < a,\\
-\frac{a^2 + 2 b r - 2 b \sqrt{ r^2-a^2 } - 2 a b\, \arcsin(a/r)}
{2\pi r a b },\quad a<r< b,\\ 
-\frac{a^2 + b^2 + r^2 - 2 b \sqrt{ r^2-a^2} - 2 a \sqrt{ r^2-b^2}}{2\pi r a b }-\\ 
\frac{ 2 a b (\arccos(b/r) - \arcsin(a/r))}{2\pi r a b}, 
\quad\quad\quad\quad  b < r <\sqrt{a^2+b^2},\\ 
0,\quad\quad\quad\quad\quad\quad\quad\quad\quad\quad\quad\quad\quad  \sqrt{a^2+b^2}< r.
\end{cases}
\endeq
%%%%%%%%%%%%%%%%%%%%%%%%%
%%%%%%%     THREAD-LIKE PHASE AND CURVE CORRELATION FUNCTION   %%%%%%%%%%
\section{Thread-like phase and curve CF}
%%%%%% \subsection{Thread-like phase and curve correlation function}
%%%%%%%%%%%%%%%%%%%%%%%%%%%%
Consider now the case where one of the sample phases  
(named again  phase 1) is formed by threads having the same normal 
section  of maximal chord $2\delta$ and area $\sigma$. Assume that $\delta$ 
be small (in a sense that will be defined later). Then, in the  limit 
$\sigma\to 0$, phase 1 shrinks to a curve $\cC$ that might  have 
branching points. The number of these points  is assumed to be negligible. 
The curve $\cC$ can be parameterized as 
\begeq\label{4.1}
\bR = \bR(\ell)=\bigl(X(\ell),\,Y(\ell),\,Z(\ell)\bigr),
\endeq 
where $\ell$ denotes  the  curvilinear coordinate of the curve point set at $\bR$.  
It is also assumed that $\cC$ is {\em smooth} in the sense that $\bR(\ell)$ 
is continuously differentiable up to the third order (included). Then, the 
curve at each of its points is endowed of two curvature radii: $R_c(\ell)$ 
and $R_t(\ell)$, respectively named {\em curvature} radius and {\em torsion} 
radius (Smirnov, 1970). To define these, one considers the three mutually 
orthogonal unit vectors defined as follows 
\begeq\label{4.2}
{\hat \tau}(\ell)\equiv \frac {d \bR(\ell)}{d\ell},\quad  
\hbn(\ell)\equiv \frac {d {\hat\tau}(\ell)}{d\ell}\Big/
\big|\frac {d {\hat\tau}(\ell)}{d\ell}\big|
\quad{\rm and}\quad {\hbb}(\ell)\equiv {\hat\tau}(\ell)\times {\hbn}(\ell),
\endeq 
and respectively named {\em tangent}, {\em normal} and {\em binormal} vectors.
Clearly, ${\hat \tau}(\ell)$ specifies the direction of the tangent to $\cC$ at 
$\bR(\ell)$, $\hbn(\ell)$ is the unit vector orthogonal to  ${\hat \tau}(\ell)$ such 
that these two vectors determine the plane osculating the curve at $\bR(\ell)$. 
Besides, the orientation of $\hbn(\ell)$ is such that the curve looks concave at the 
considered point. The expressions of the aforesaid radii are as follows 
\begeq\label{4.3}
R_c(\ell)\equiv 1\Big/\Big|\frac {d {\hat\tau}(\ell)}{d\ell}\Big| \quad{\rm and}
\quad R_t(\ell)\equiv 1\Big/\Bigl(\frac{d\hbb(\ell)}{d\ell}\cdot\hbn(\ell)\Bigr).
\endeq 
The curvature radius is the radius of the circumference osculating the curve while 
the torsion radius specifies how the osculating plane rotates as one moves 
along the curve. It follows that the curvature radius is always positive while the 
torsion radius can be positive at some points of the curve and negative at others.\\ 
Similarly to equation (\ref{3.1}) one defines the mean of the curvature radii of 
the curve as
\begeq\label{4.7}
\bRc\equiv{\langle \big(1/{R_c}^2+1/{R_t}^2\big)\,\rangle_{\cC}}^{-1/2},
\endeq 
where the average  is performed integrating all over the length of $\cC$. 
(It is noted that $\cC$ generally consists of many disjoint closed and/or open 
curves.) \\ 
One goes now back to $P_{1,1}(r)$, the SPF of the thread-like 
phase, to analyze in more detail the limit $\sigma\to 0$.  First, it is observed 
that $\sigma$ is small if $\delta<< \bRc$ as it will hereafter be assumed. Let 
$\br$ be the position vector of a point of phase 1 and consider the plane that 
contains this point and is orthogonal to $\cC$. Denote respectively by $\bR(\ell)$ and by 
$\usig(\ell)$ the intersections of the plane with $\cC$ and with the 
thread-like phase. One can then write $\br=\bR(\ell)+\vxip$ where 
$\vxip$ lies within $\usig(\ell)$, the normal section set of the thread at $\bR(\ell)$.    
 %%\end{document}
Function $P_{1,1}(r)$ can be written as 
\begin{eqnarray}
&&P_{1,1}(r) =\frac{1}{4\pi\, V}
\int_{\cC}\rd\ell_1\int_{\usig(\ell_1)} d^2\vxipu\,\int_{\cC}\rd\ell_2
\int_{\usig(\ell_2)} \rd^2\vxipt \times  \label{4.9}  \\
&&\quad\quad\quad\quad\quad\quad  \int \rd\hw\,
\delta_3\big(\bR(\ell_1)+\vxipu+r\hw-\bR(\ell_2)-\vxipt\big).       \nonumber       
\end{eqnarray}                                                                             %%\end{document}    
The volume fraction of the thread-like phase is fairly approximated by 
$\phi_1=\sigma L/V$ with $L$ equal to the total length of the threads.  From 
equation (\ref{2.3a}) follows that 
\begeq\label{4.8}
P_{1,1}(0)=\sigma L/V,\quad{\rm and}\quad P_{1,1}(\infty)=\sigma^2 (L/V)^2.
\endeq 
Similarly to the case discussed in \S\,3.1, in the range $0<r<2\delta$, 
$P_{1,1}(r)$ is fairly approximated by equation (\ref{3.5}).  This can be written in the 
algebraic form
\begeq\label{4.10}
P_{1,1}(r)\approx {\mathfrak P}_{1,\cC}(r)\equiv
\frac{\sigma\,L}{V}- \frac{r\,\pi\delta\,L}{2V} + \frac{\pi\,r^3\,L}
{64\,V\,\delta},\quad\quad r<2\delta,
\endeq 
if one approximates the  
threads by circular cylinders  so that $S\approx 2\pi\delta L$, 
$H\approx\frac{1}{2\delta}$ and $K_G\approx 0$.\\ 
In the range $r>2\delta$, owing to the smallness of $\delta$, one can neglect 
the dependence on $\vxipu$and $\vxipt$ within the Dirac function. Then,   
the integrals over these variables can explicitly be performed and  one finds that 
\begeq\label{4.12}
P_{1,1}(r)\approx \frac{\sigma^2\,L}{V}\gamma_{\cC}(r),\quad\quad r>2\delta,
\endeq 
with  
\begeq\label{4.11}
\gamma_{\cC}(r)\equiv \frac{1}{4\pi L} \int \rd\hw\,\int_{\cC}d\ell_1\int_{\cC}\rd\ell_2\,
\delta_3\bigl(\bR(\ell_1)+r\hw-\bR(\ell_2)\bigr), \quad r>0.
\endeq 
One concludes that the $r$-dependence of the SPF of a thread-like phase   is 
described by function $\gamma_{\cC}(r)$ in  the outer distance range and by 
equation (\ref{4.10}) in the inner one. \\ 
Function $\gamma_{\cC}(r)$ is determined by curve $\cC$ and  will 
be named {\em curve correlation function}. The dimensions of $\gamma_{\cC}(r)$ 
are ${\rm length}^{-2}$.  \\
 First one elaborates equation  (\ref{4.11}) so as to convert it into a one dimensional 
integral. To this aim,  the equation is written as  
\begeq\label{4.14}
\gamma_{\cC}(r)=\frac{1}{4\pi\, L\,r^2}\,\int_{\cC}\rd\ell_1\int_{\Sigma(\bR(\ell_1),r)}
dS\int_{\cC}\rd\ell_2\,
\delta\bigl(\bR(\ell_1)+\bR-\bR(\ell_2)\bigr),
\endeq 
where $\bR$  denotes the position vector of the infinitesimal surface  
element $dS$  of $\Sigma(\bR(\ell_1),r)$, the sphere of radius $r$ centered at 
the infinitesimal curve element $d\ell_1$ set at $\bR(\ell_1)$. 
Note that the modulus of $\bR$ is $r$. One denotes the function defined by the 
innermost two integrals of (\ref{4.14}) as 
\begeq\label{4.15}
p_{\cC}(\ell_1,r)\equiv \int_{\Sigma(\bR(\ell_1),r)}\rd S\int_{\cC}\rd\ell_2\,
\delta\bigl(\bR(\ell_1)+\bR-\bR(\ell_2)\bigr). 
\endeq
The function can easily be evaluated choosing a Cartesian orthogonal frame having 
the origin $O$ at the position of $d\ell_1$ so that $\bR(\ell_1)={\bf 0}$. The value 
of the integral can be different from zero if and only if $\Sigma({\bf 0},r)$ 
intersects $\cC$. Assume that the intersection points exist and let $P$ denotes one
fo  these points.  Then, one chooses the $Z$ axis along $OP$. The plane containing  
$dS$ is perpendicular to $Z$ while $d\ell_2$ forms an angle, denoted by $\theta$, 
with axis  $Z$, so that 
$dZ=\cos(\theta)d\ell_2$. Around $P$, $p_{\cC}(\ell_1,r)$ takes the form  
$\int \rd X\int \rd Y\int (\rd Z/cos\theta)\delta_1(X)\delta_1(Y)\delta_1(Z)$ and  
its value is $1/\cos (\theta)$. 
Recalling that $\rd\ell_2$ is parallel to ${\hat\tau}(\ell_2)$, with ${\hat\tau}$ 
defined by equation (\ref{4.2}a),  one gets  
\begeq\label{4.16}
\hw(\ell_1,r)=(\bR({\overline\ell}_2)-\bR(\ell_1))/r,
\endeq
\begeq\label{4.17}
\cos\big(\theta(\ell_1,r)\big)=\hw(\ell_1,r)\cdot{\hat\tau}({\overline\ell}_2), 
\endeq 
and 
\begeq\label{4.17b}
p_{\cC}(\ell_1,r) =\sum_{i} \frac{1}{\hw_i(\ell_1,r)\cdot{\hat\tau}_i({\overline\ell}_2)}=
\sum_{i} \frac{1}{\cos\big(\theta_i(\ell_1,r)\big)}, 
\endeq
where index $i$ labels all the points of $\cC$ that are at distance $r$ from the 
point set at $\bR(\ell_1)$. It is stressed that  equation (\ref{4.16})  determines 
the direction of $\hw$ as well as the curvilinear coordinate value $\ell_2$ 
(denoted by  ${\overline\ell}_2$) owing to the fact  that the vector on the right 
side of (\ref{4.16}) must be unimodular. By (\ref{4.14}) and (\ref{4.17b}) one 
concludes that the curve CF is proportional to the curvilinear average of 
$1/\cos\theta(\ell_1,r)$, \ie
\begeq\label{4.18}
\gamma_{\cC}(r)=\frac{1}{4\pi \,r^2} \big\langle \frac{1}{\cos\theta(.,r) }\big\rangle_{\cC}.
\endeq 
The leading asymptotic behavior of the above expression  as  $r\to 0$ is worked
 out in appendix  C and reads 
\begeq\label{4.19}
\gamma_{\cC}(r)\approx \frac{1}{2\pi \,r^2} + 
\frac{1}{16\pi}\big\langle \frac{1}{{R_C}^{2}}\big\rangle_{\cC} +o(r^2) ,
\quad\quad r\approx 0.
\endeq 
This result  shows that the curve CF diverges as $1/r^2$ as one approaches 
the origin. The leading behavior of the curve CF as $r\to\infty$ is immediately 
obtained by equations (\ref{4.12}) and (\ref{4.8}b) and reads 
\begeq\label{4.19a}
\gamma_{\cC}(r)\approx L/V +O(1/r),\quad\quad r\to \infty.
\endeq 
The moments $M_{\cC,m}$ of the curve CF, similarly to the case of the surface CF, can be 
expressed in terms of the moments of the curve. In fact 
\begeq\label{4.19ax}
M_{\cC,m}\equiv\int_0^{\infty}r^{2m+2}\gamma_{\cC}(r)dr=
\frac{1}{4\pi L}\int_{\cC}d\ell_1\int_{\cC}d\ell_2\bigl(\bR(\ell_2)-\bR(\ell_1)\bigr)^{2m}.
\endeq 
From this relation follows that $M_{\cC,0}=L/4\pi$.
%%%%%          EXAMPLES OF CURVE CFs       %%%%%%%%%%%
 \subsection{Examples of curve  CFs}     %%%%%%%%%%%%%
Two examples of  curve CFs are  reported in the following subsections.
%%%%%%%%%%     \subsubsection{The LINEAR SEGMENT}   %%%%%%%%%%%%
\subsubsection{The curve CF of a linear segment}
The first refers to a linear segment of length $L$. Applying definition 
(\ref{4.14}), the angle defined by (\ref{4.17}) is equal to zero. Thus, 
equation (\ref{4.18})  yields
\begeq\label{4.20}
\gamma_{\cC,ls}(r,L)=
\begin{cases}
\frac{1}{2\,\pi\,r^2}-\frac{1}{2\,\pi\, L \,r}&\text{if}\quad 0<r<L,\\
0&\text{if}\quad L<r.
\end{cases}
\endeq  
%%%%%%%%%%%%%%%%%%%%%%%%%%%%%%%%%%%%%%%%%%%%%%%%%%%%%%%%%%
\begin{figure}[!h]                   %%       FIGURE 3       NEEDLE-case         %%%%%%%%%%%%%    
{\includegraphics[width=7.truecm]{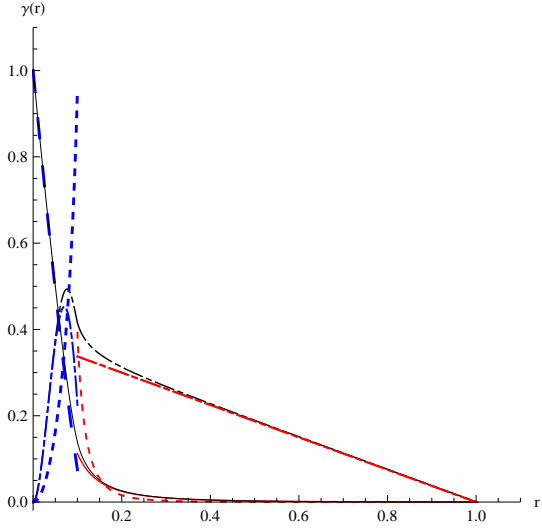}}
%% /Users/salvino/Desktop/WORK_IN_PRGS/THREAD_FILM_LIKE _SYSTEMS/cylinder_NEW_a
\caption{\label{Fig3} {The  curves have the same meaning as in  
Fig.\,\ref{Fig2}. They do however refer to a circular cylinder of radius $R=\frac{1}{20}$u 
and height $h=1$u.  The black, blue and red dash-dotted curves are the plots of 
$300\cdot r^2\gamma_{ndl}(r)$ [see Eq. (\ref{A.6})],   $300\cdot r^2{\mathfrak P}_{1,c}(r))$ 
[see Eq. (\ref{4.10})] and $300\cdot  r^2 h\gamma_{\cC,ls}(r)$ [see Eq. (\ref{4.20})].  The dotted blue 
and red curves are the 
plot of the error, multiplied by 15, in $0<r<2R$ and $r>2R$, respectively. }}    
\end{figure}
By letting the radius of a circular cylinder go to zero, the limit 
of its CF [multiplied by $V/L\sigma^2$  as required by 
(\ref{4.12})] must reproduce (\ref{4.20}). One easily checks that 
this property holds true evaluating the aforesaid limit stemming 
from $\gamma_{ndl}(r,R,h)$, the cylinder CF given by equation  (\ref{A.6}). 
Besides this CF can be approximated by (\ref{4.10}) if $r<2R$ and 
by $(\pi R^2)\gamma_{\cC,ls}(r,h)$ if $r>2R$. 
Fig.\,\ref{Fig3} shows the achieved accuracy in the case $R=\frac{1}{20}$u and $h=1$u. 
%%%%           The curve CF of a circumference}
\subsubsection{The curve CF of a circumference}
The curve CF of a circumference of radius $R$ also can easily be  evaluated. 
One  puts $\ell=R\varphi$. Given $r$, the associated $\varphi$ value is 
$\varphi=\arccos\big((2R^2-r^2)/2R^2\big)$.  Then 
one finds that  $\theta(\ell,r)=\pi/2-(\pi-\varphi)/2$ so that 
$\cos\theta(\ell,r)=\frac{\sqrt{4R^2-r^2}}{2R}$. Finally, from 
(\ref{4.18}), one gets the circumference CF
\begeq\label{4.21}
\gamma_{\cC,crc}(r)=
\begin{cases}
\frac{R}{\pi\,r^2\sqrt{4R^2-r^2}}& \text{if}\quad 0 < r < 2R,\\
0 & \text{if}\quad 2R<r.
\end{cases}
\endeq
This CF behaves as ${1}/{2\pi\,r^2}$ as $r\to 0$ and diverges as 
$1/[2\pi R^{3/2}(2R-r)^{1/2}]$ as  $r\to 2R^-$.
%%%%%%%%
%%%%       SCATTERING INTENSITY BEHAVIOUR  %%%%%%%%%%%%
%%%%%%%%%%
\section{Scattering intensity behavior}
The scattering intensity $I(q)$ is the Fourier transform (FT) of equation (\ref{2.5}) 
times $V\langle\eta^2\rangle$ and is, therefore, equal to 
\begin{eqnarray}\label{5.1}
I(q)&&=\frac{4\,\pi\,V\langle\eta^2\rangle}{q}\int_0^{\infty}r\,
\sin(q\,r)\gamma(r)\rd r.
\end{eqnarray}
This can be recast into the form (Ciccariello\& Riello, 2007)
\begeq\label{5.1x}
I(q)=  \frac{V}{2}\sum_{i=1}^3{n(i;j,k)\phi_i(1-\phi_i)}{\tilde\Gamma}_i(q)
\endeq
with
\begeq\label{5.2}
{\tilde\Gamma}_i(q)\equiv \frac{4\,\pi}{q}\int_0^{\infty}r\,\sin(q\,r)\Gamma_i(r)\rd r.
\endeq 
The existence of the integral is ensured by condition  (\ref{2.7}b), while equation 
(\ref{2.7}a) implies the following sum-rule 
\begeq\label{5.2a}  
\int_0^{\infty}q^2\,{\tilde\Gamma}_i(q)\,\rd q = 2\,\pi^2,
\endeq
responsible for the Porod invariant relation
\begeq\label{5.3}  
\cQ_{P}\equiv\int_0^{\infty}q^2\,I(q)(q)\,\rd q = 2\,\pi^2\langle\eta^2\rangle.
\endeq
Besides,  the ${\tilde\Gamma}_i(q)$s  obey the condition 
${\tilde\Gamma}_i(q)\ge 0$ whatever  the scattering vector $q$ because 
they are the FTs of  the $\Gamma_i(r)$s that have a convolution structure 
owing to definitions (\ref{2.4}) and (\ref{2.3}).  However, these  conditions 
are not sufficient, on the basis of (\ref{5.1}), to ensure the required 
positiveness of $I(q)$ because some of the $n(i;j,k)$s can be negative. 
Hence, the last quantities have to obey appropriate constraints to ensure 
the positiveness of $I(q)$. This conclusion, that might look at first 
surprising, is a consequence of the functional density theorem that 
states that the correlation function is uniquely determined by the 
density value and the interaction potential. On the basis of this 
statement it is clear that the assignment of the sample internal geometry, 
equivalent to assigning  the ${\tilde\Gamma}_i(q)$s, cannot be fully 
independent of the $n_i$ values.\\ 
Each $P_{i,i}(r)$ is determined by the length distribution of the 
chords  that have  both ends within the related phase. In most of 
the cases the distribution is not uniform but has a particular shape that 
naturally defines some length values $L_1,\ldots,L_M$  as the bounds 
of the ranges relevant to  the small and the large distances asymptotic 
behaviors of $P_{i,i}(r)$ as well as to the positions of possible peaks and 
shoulders.  Since $\Gamma_i(r)$ is obtained from $P_{i,i}(r)$ by sutracting 
to this quantity $\phi_i^2$ it happens that $\Gamma_i(r)$  is certainly 
positive around the origin  owing to (\ref{2.7}a) and that it can be negative  
for other $r$ values.     The aforesaid  distances 
$L_1<\ldots<L_M$ in turns define some particular scattering vector 
values ${\bar q}_i$ $(i=1,\ldots)$  through the well known relation 
${\bar q}_i=2\pi/L_i$. The ${\bar q}_i$s are the reciprocal space values 
where one expects a change in the ${\tilde\Gamma}_i(q)$ behavior. \\ 
One applies now these considerations to the film-like and thread-like 
systems discussed in the previous sections. There it was shown that the 
$P_{1,1}(r)$ SPF is certainly endowed of  two typical lengths: $\bRs$ 
[$\bRc$] [see equations (\ref{3.1}) and (\ref{4.7})] and $\delta$ 
$[2\delta]$ with $\bRs>\delta$  [$\bRc>2\delta$] and  that  
$P_{1,1}(r)$ is fairly approximated by a third degree polynomial if 
$r<\delta[2\delta]$ and by  a surface [a curve] CF if $r>\delta$ 
$[r>2\delta]$.  It is convenient to separately  discuss the case of  
film-like systems and  that of thread-like ones. 
%%%%        INTENSITY BEHAVIOUR    IN THE FILM-LIKE CASE      %%%%
%%%%%%     Intensity behavior in the  film-like case       %%%%%%%%%%
\subsection{Intensity behavior in the  film-like case}
To begin with one analyzes first the behavior of ${\tilde\Gamma}_1(q)$ 
under the further simplifying assumption that one only has two typical 
lengths,\ie\,$L_1=\delta$ and $L_2=\bRs$. Thus, the 
${\tilde\Gamma}_1(q)$ behavior is expected to change as one passes
 from the $q$-range $[0,\,2\pi/\bRs]$ to  $[2\pi/\bRs,\, 2\pi/\delta]$ 
and, finally, to $[ 2\pi/\delta,\,\infty]$. By  (\ref{2.4}) and the results 
of \S\,3 one finds that 
\begeq\label{5.3a}
\Gamma_1(r)\approx \begin{cases}{{\bar\GP}_1(r)}
\quad\text{if}\quad 0<r<\delta\\
{{\bar \gamma_{\cS}}(r)}\quad\text{if}\quad \delta<r,
\end{cases}
\endeq  
where it has been put
\begeq\label{5.3b}
{\bar\GP}_1(r)\equiv \frac{{\GP}_1(r)-\phi_1^2}{\phi_1(1-\phi_1)},
\quad{\rm and}\quad {\bar \gamma}_{\cS}(r)\equiv 
\frac{\delta^2\,S_f\,\gamma_{\cS}(r)/V-\phi_1^2}{\phi_1(1-\phi_1)}.
\endeq  
The FT of equation  (\ref{5.3a}) yields
\begeq\label{5.3c}
{\tilde\Gamma}_1(q)={\tilde{\bar\GP}}_1(q)+
{\tilde{\bar \gamma}}_{\cS}(q),
\endeq 
with 
\begeq\label{5.3ca}
{\tilde{\bar\GP}}_1(q)=
\frac{4\pi}{q}\int_0^{\delta}r\sin(q\,r){\bar\GP}_1(r)\rd r
\endeq 
and 
\begeq\label{5.3d}
{\tilde{\bar\gamma}}_{\cS}(q)=
\frac{4\pi}{q}\int_{\delta}^{\infty}r\sin(q\,r){\bar\gamma}_{\cS}(r)\rd r.
\endeq 
[It is observed that condition (\ref{3.9}) ensures that 
${\bar \gamma_{\cS}}(r)\to 0$ as 
$r\to\infty$ so that  integral (\ref{5.3d}) exists.]  The behavior of ${\tilde\Gamma}_1(q)$ 
within the first $q$-range $[0,\,2\pi/\bRs]$ could be obtained by expanding the 
function around $q=0$. One  finds
\begeq\label{5.3e}
{\tilde\Gamma}_1(q)\approx \sum_{n=0}^N \frac{(-1)^n}{(2\,n+1)!}
\big(M_{\GP,2n}+M_{\gamma_{\cS},2n}\Bigr)\,q^{2n}+o(q^{2N})
\endeq 
with  
\begeq\label{5.3f}
M_{\GP,2n}\equiv\int_0^{\delta}r^{2n+3}{\bar{\GP}}_1(r)\rd r 
\endeq
and
\begeq\label{5.3g}
M_{\gamma_{\cS},2n}\equiv\int_{\delta}^{\infty}r^{2n+3}
{\bar\gamma}_{\cS}(r)\rd r. 
\endeq
One sees that $M_{2n}\,[\equiv (M_{\GP,2n}+M_{\gamma_{\cS},2n})]$ is 
the $2n$th moment of $\Gamma_1(r)$. Since $M_{\GP,2n}$ is 
$O(\delta^{2n+3})$, the value of $M_{2n}$ is essentially equal to the 
corresponding momentum of ${{\bar\gamma}_{\cS}(r)}$ and strongly 
depends on the way  ${{\bar\gamma}_{\cS}(r)}$ approaches zero as 
$r\to \infty$. Since the last  behavior generally is not known, on a general 
ground the only existence and positiveness of $M_0$ is certain. The 
existence of the other higher moments is certain only if one  assumes that 
${{\bar\gamma}_{\cS}(r)}$ approaches zero by an exponential decrease 
or, more strongly, that the sample is a dilute collection of particles with 
a finite maximal size and that the inter-particle interference may be 
 neglected (Guinier \& Fourn\'et, 1955). \\
Consider now the second $q$-interval $[2\pi/\bRs,\,2\pi/\delta]$. If $q$ is 
close to $2\pi/\bRs$, contribution ${\tilde{\bar\GP}}_1(q)$ to 
${\tilde\Gamma}_1(q)$  can still, though less accurately, be approximated 
by the sum $(M_{\GP,0}-M_{\GP,2}q^2/6+\ldots)$ 
[see (\ref{5.3e})]. To estimate  contribution ${\tilde{\bar \gamma}}_{\cS}(q)$ 
one must determine its asymptotic behavior as $q\to\infty$. Integrating 
equation (\ref{5.3d}) by parts  one finds that the leading term is 
\begeq\nonumber
{\tilde{\bar \gamma}}_{\cS}(q)\approx \frac{4\pi}{q^2}
\frac{\delta^3S_f\gamma_{\cS}(\delta)/V-\delta\phi_1^2}{\phi_1(1-\phi_1)}
\cos(q\delta). 
\endeq
In this expression, the contribution related to $(-\delta\phi_1^2)$ is 
canceled by the corresponding contribution present in 
${\tilde{\bar\GP}}_1(q)$ since ${{\bar\GP}}_1(r)$ also presents the 
constant term $-\phi_1^2$. Thus one can write 
\begeq\label{5.3hx}
{\tilde{\bar \gamma}}_{\cS}(q)\approx \frac{4\pi}{q^2}
\frac{\delta^3S_f\gamma_{\cS}(\delta)/V}{\phi_1(1-\phi_1)}\cos(q\delta). 
\endeq
provides one also writes 
\begeq\label{5.3ia}
{\tilde{\bar\GP}}_1(q)\approx (M_{\GP,0}\p-M_{\GP,2}\p q^2/6)+\ldots
\endeq
where the primes denote that the  moments have been 
evaluated by (\ref{5.3e}) without subtracting $\phi_1^2$ from ${\GP}_1(r)$. 
In the sub-interval of the considered $q$-range such that $q\bRs>2\pi$ and 
$q\delta<1$, using (\ref{3.8}) and the relation $\delta S_f/V\approx\phi_1$,   
from (\ref{5.3hx})  one gets
\begeq\label{5.3i}
{\tilde{\bar \gamma}}_{\cS}(q)\approx \frac{2\pi}{q^2}
\frac{\delta}{1-\phi_1}. 
\endeq
Recalling that $M_{\GP,0}\p$ is $O(\delta^3)$ one finds that 
$M_{\GP,0}\p\propto\delta (q\delta)^2/q^2$ that, compared to the above 
contribution, can be neglected. One concludes that 
\begeq\label{5.3j}
{\tilde\Gamma}_1(q)\approx  \frac{2\pi}{q^2}\frac{\delta}{1-\phi_1}\  \ 
{\rm if}\ \ \frac{2\pi}{\bRs}<q<\frac{1}{\delta}.
\endeq
Once one considers the third $q$-range, \ie\,$\frac{2\pi}{\delta} <q$,  in direct 
space the sample structure is analyzed on a distance scale smaller that $\delta$. 
Then, the limit $\delta\to 0$ is no longer valid and the large $q$ behavior of 
${\tilde\Gamma}_1(q)$ must be determined starting from the FT of 
${\Gamma}_1(r)$ defined by (\ref{2.4}). An integration by parts and the use of 
(\ref{2.3c}) immediately yields the equivalent of the Porod relation
\begeq\label{5.1.3}
{\tilde\Gamma}_{1}(q)\approx  \frac{4\,\pi\,S_f}{V\,\phi_1(1-\phi_1)\,q^4}\quad 
{\rm if}\quad 
\frac{2\pi}{\delta}<q.
\endeq        %%%%%\end{document}
Confining oneself  to three phase film-like systems where  film-like phase 1 
fully separates phase 2 from phase 3 (so that the last two phases have no common 
interface), phases 2 and 3 are characterized by the only length $\bRs$.  
Then, by the same  considerations that yielded (\ref{5.1.3}), one concludes that the 
asymptotic leading terms of ${\tilde\Gamma}_{2}(q)$ and ${\tilde\Gamma}_{3}(q)$ 
are  
\begeq\label{5.1.4}
{\tilde\Gamma}_{i}(q)\approx\begin{cases}
O(\delta^3)\ \ &\text{if}\ \ q<\frac{2\pi}{\bRs}\ \ \ i=2,3,\\
   \frac{2\,\pi\,S_f}{V\,\phi_i(1-\phi_i)\,q^4},\ \ 
 &\text{if}\  \  \frac{2\pi}{\bRs}<q\ \ \ \ i=2,3.
\end{cases}
\endeq          %%%%%%\end{document}
Substituting the above asymptotic behaviors within equation (\ref{5.1}) it results that 
the asymptotic leading term of the scattering intensity relevant to a three-phase
 film-like system (with no common interface between phases 2 and 3) is 
\begeq\label{5.1.5}
I(q)\approx 
\begin{cases} \frac{\cP_{\cS}}{q^2}\ &\text{ if } \quad 
\frac{2\pi}{\bRs}<q<\frac{1}{\delta},  \\
 \frac{\cP}{q^4}\ &\text{ if } \quad 
\frac{2\pi}{\delta}<q.
\end{cases}
\endeq             
with 
\begin{eqnarray}\label{5.1.5a}
\cP_{\cS}&\equiv &{\pi\,n(1;2,3)\,\delta^2\,S_f},\\
\cP\ &\equiv &2\, \pi\,S_f\,\big((n_1-n_2)^2+(n_1-n_3)^2\big).\label{5.1.5ab}
\end{eqnarray}  %%  \end{document}
The main conclusion of this analysis is that: {\em if $\delta<<\bRs$, the {\em log-log} 
plot of the scattering intensity of a three-phase film-like system shows a linear 
behavior with slope -2 at intermediate $q$-values and a further linear 
behavior with slope -4 in the outer $q$-range}  (of course, provided $q_{max}$, 
the largest observed scattering vector, obeys to $q_{max}\delta>>2\pi$). 
Hence,  the lower bounds $q_s$ and $q_l$ of the two linear behavior ranges 
yield an estimate of $\bRs$ and $\delta$ through the relations 
$\bRs\approx 2\pi/q_s$ and $\delta\approx 2\pi/q_l$. The intersections of 
the two straight lines with the vertical axis set at $\log( q)=0$ determine 
the values of $\log(\cP_{\cS})$ and $\log(\cP)$. 
According to (\ref{5.1.5a}) and (\ref{5.1.5ab}), the resulting ratio $\cP_{\cS}/\cP$ 
determines $\delta$ if the phase contrasts are known and then, either 
$\cP_{\cS}$ or $\cP$ can be used to determine the film area $S_f$.\\
The assumption that phases 2 and 3 have no common 
interface is now relaxed. Then, equations (\ref{5.3j}) and  (\ref{5.1.3})  
remain unchanged while the second of equations (\ref{5.1.4}) converts into 
\begeq\label{5.1.5b}
{\tilde\Gamma}_{2}(q)\approx   
\frac{2\,\pi(S_{1,2}+S_{2,3})}{V\,\phi_2(1-\phi_2)\,q^4},
\ \ \ {\rm  }\  \  \frac{2\pi}{\bRs}<q
\endeq 
and a similar one for ${\tilde\Gamma}_{3}(q)$ with $S_{1,2}+S_{1,3}=S_f$. 
From these relations and (\ref{5.1}) one obtains the general 
expressions of $\cP_{S}$ and $\cP$
\begin{eqnarray}\label{5.1.5ax}
\cP_{\cS}&\equiv &{\pi\,n(1;2,3)\,\delta^2\,(S_{1,2}+S_{1,3})},\\
\cP\ &\equiv &2\, \pi\big[S_{1,2}(n_1-n_2)^2+S_{1,3}(n_1-n_3)^2+
S_{2,3}(n_2-n_3)^2\big].\label{5.1.5abx}
\end{eqnarray} 
Though these two relations, even combined with the Porod invariant one, are no longer sufficient 
to determine the four structural parameters $\delta,\, S_{1,2},\, S_{1,3}$ and $S_{2,3}$,  they yield 
however useful general bounds on these parameters. \\
%%%%   \vskip 2truecm \end{document}
Two illustrations are now reported. Although they are na\"ive mathematical 
models  they illustrate the main points of the previous discussion.  The first 
illustration deals  with the scattering intensity of a  spherical shell whose  
\begin{figure}[!h]                   %%       FIGURE                    %%%%%%%%%%%%%    
%%  SPERICALSHELL    see "Spherical_Shell.nb"
{\includegraphics[width=7.truecm]{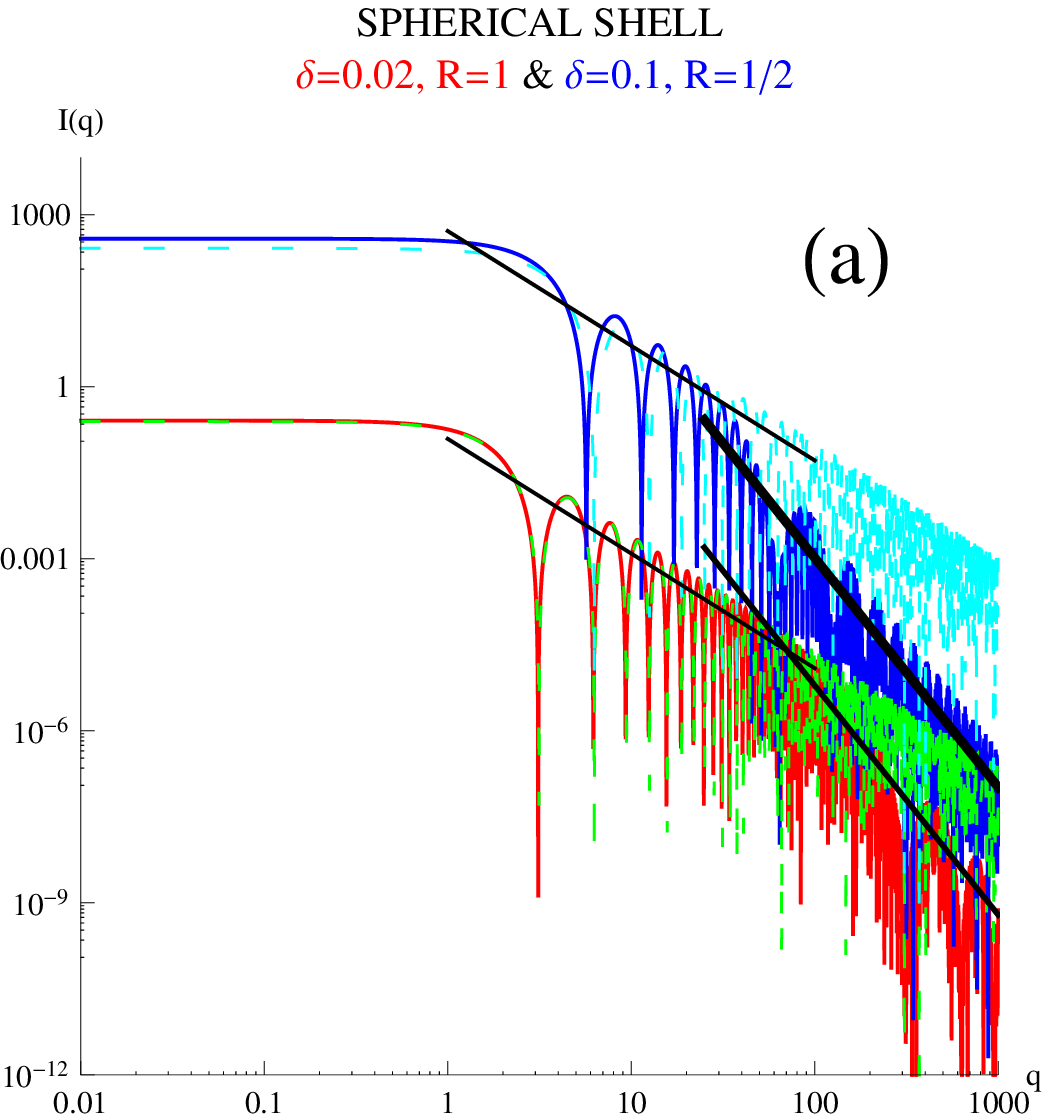}}
{\includegraphics[width=7.truecm]{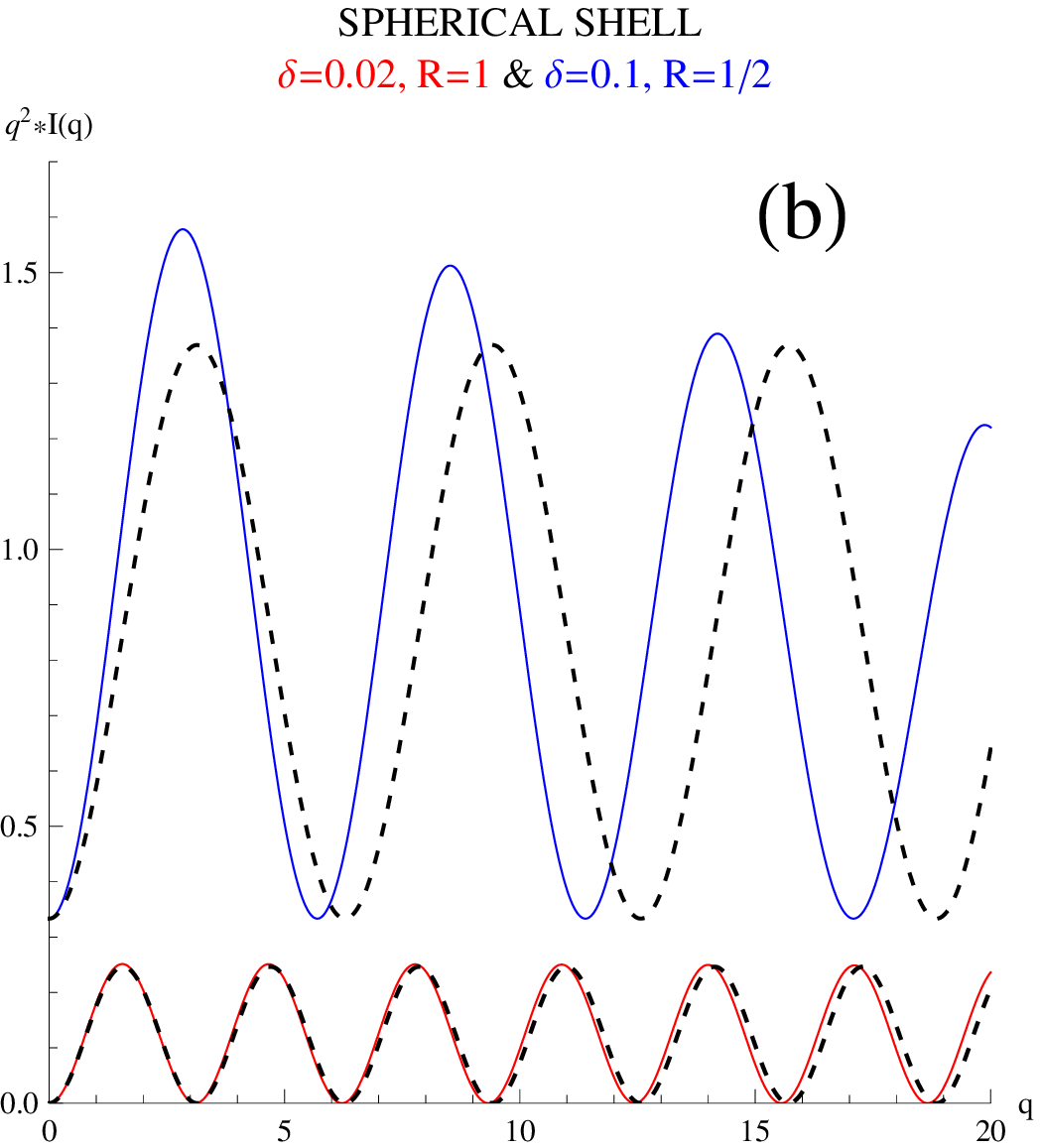}} 
\caption{\label{Fig4} {(a) The red and the blue curve are the plots of the FTs 
of equation  (\ref{3.25}) with $(R,\,\delta)$ respectively equal to (1,\,0.02)u 
and (0.5,\,0.1)u, while the  green and the cyan curves are the FTs of  
(\ref{3.26}). The blue and the cyan intensities have been multiplied by $10^3$  
for greater clarity.  (b) Kratky plots of  
the previous intensities. The dotted curves refer to the FTs of  (\ref{3.26}).  
The blue curves have  vertically been shifted by 0.3. The agreement improves  
as $\delta/R$ gets  smaller. [The $q$ units are u$^{-1}$.]
%%%%   SEE THE FILE  %% ""/Users/salvino/Desktop/WORK_IN_PRGS/THREAD_FILM_LIKE_SYSTEMS"/FTs_cylinder"
}}
\end{figure}
$(R,\,\delta)$parameters take the values: (1,\,0.02)u and (0.5,\,0.1)u. 
The left panel of Fig. \ref{Fig4} shows, in red and blue, the spherical shells'  
scattering intensities obtained by the FT of  (\ref{3.25}) and, in cyan and green, the 
corresponding FTs of $\frac{\delta^2 S_1}{V}\gamma_{\cS,sph}(r,R)$ [see  (\ref{3.26})], 
equal to  
\begeq\label{5.1.6}
\frac{\delta^2 S_1}{V}{\tilde\gamma}_{\cS,sph}(q)=12 \pi R^2 \delta\, \sin^2(q\, R)/
[q^2 (3 R^2 + 3 R \delta + \delta^2)].
\endeq 
[This FT has been obtained by integrating over $[0,\,2R]$, the total support  of  $\gamma_{\cS,sph}(r,R)$.]
The thin and the thick continuous straight lines are the plots of the leading 
asymptotic terms given by equation  (\ref{5.1.5}) [and  (\ref{5.1.5a}) and 
(\ref{5.1.5ab})]. The figure shows that the intensity relevant to the spherical 
shell [leaving aside the oscillations] shows both a $q^{-2}$ and a 
$q^{-4}$ behavior and that the intensity of the associated spherical 
surface well reproduces the first intensity throughout the range $[0,\,2\pi/\delta]$ 
(\ie\,not only within $[2\pi/\bRs,\,2\pi/\delta]$). Moreover it results that 
the smaller the ratio $\delta/2R$ the wider becomes the range where the $q^{-2}$ 
behavior occurs. The right panel of Fig. \ref{Fig4} shows the $q^2I(q)$ versus $q$
 plots (also known as Kratky plots) of the two considered spherical shells 
(continuous curves) and of their associated spherical surfaces (broken curves). 
The agreement is much better in the case (1,\,0.02) owing to  the smaller 
$\delta/2R$  value.\\ 
 \begin{figure}[!h]                   %%       FIGURE                    %%%%%%%%%%%%%    
{\includegraphics[width=7.truecm]{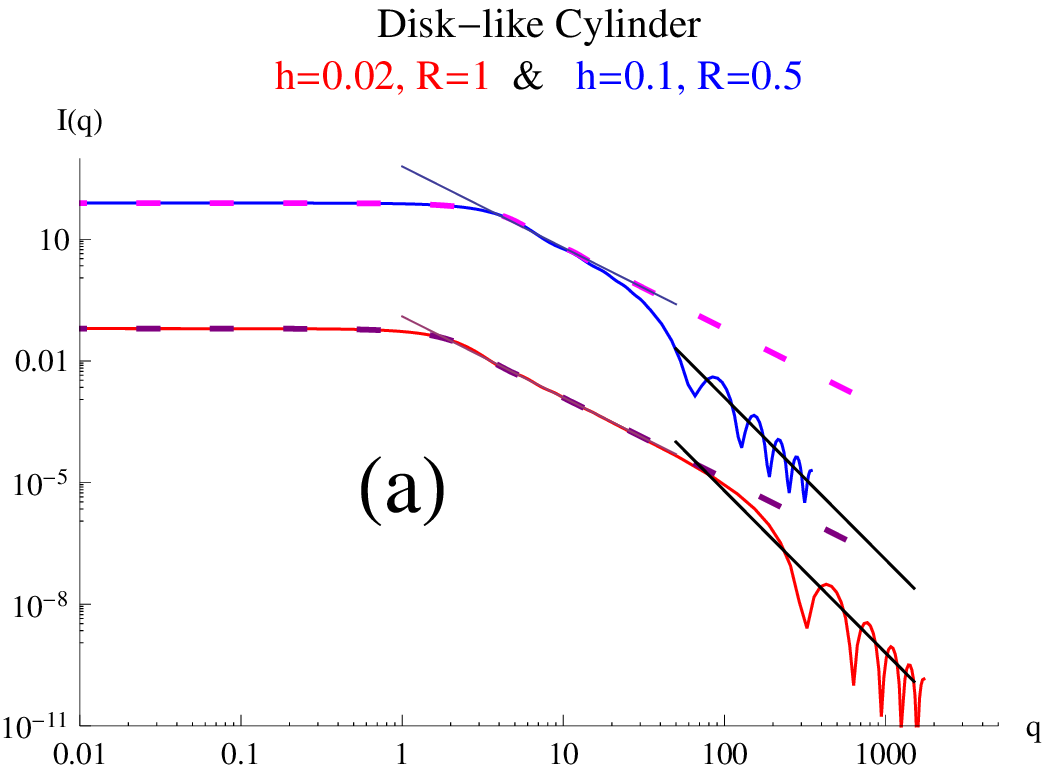}}
{\includegraphics[width=7.truecm]{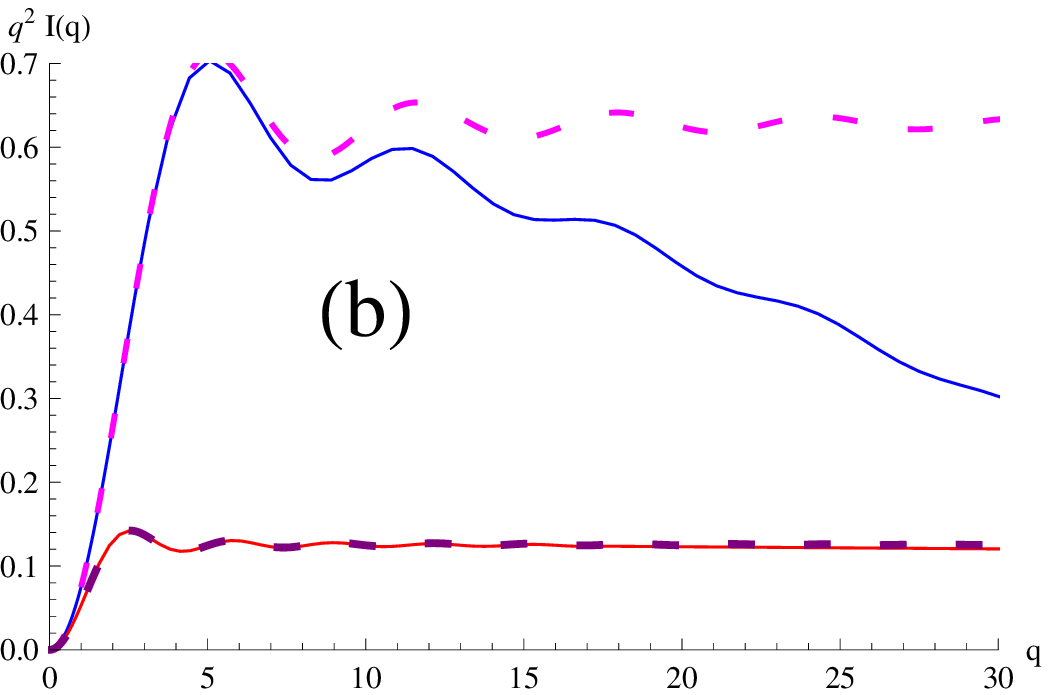}}
\caption{\label{Fig5} {(a): The red and the blue curve are the plots of the FTs of equation 
 (\ref{A.5}) with $(R,\,h)$ respectively equal to (1,\,0.02)u and (0.5,\,0.1)u, while 
the  magenta and the purple curves are the associated FTs of  (\ref{3.26}). 
The blue and the magenta intensities have been multiplied by $10^3$  for greater 
clarity;  (b): Kratky plots of the previous intensities (without 
any scaling factor). The broken curves refer to equation (\ref{5.1.7}). }}
%%%%   SEE THE FILE  %% ""/Users/salvino/Desktop/WORK_IN_PRGS/THREAD_FILM_LIKE_SYSTEMS"/FTs_cylinder"
\end{figure}  
The second illustration deals with the case of two disk-like cylinders characterized by 
$(h,R)=(0.02,1)$u  and  $(h,R)=(0.1,0.5)$u. 
The corresponding intensities are obtained by numerically Fourier transforming 
equation (\ref{A.5}) and respectively yield the red and the blue continuous curves 
shown in Fig. \ref{Fig5}a. The associated surface-like intensities are the 3D  FTs of 
(\ref{3.27}) multiplied by $h^2 S_f/V$, \ie
\begeq\label{5.1.7}
\frac{h^2 S_f}{V}\,{\tilde\gamma}_{\cS,crcl}(q)=\frac{2\, h\, \pi}{q^2}
\Big [1 -  {\rm J}_0( 2 q R) -{\rm J}_2( 2 q R)\Big].
\endeq
Their plots are shown as magenta and purple broken curves. They practically 
superpose to the red and blue curves throughout $[0,\,2\pi/\delta]$. 
The black lines are the plots of the relevant leading asymptotic behaviors given 
by (\ref{5.1.5}). One again observes that the 
$q$-range, where the intensity behavior is $1/q^2$ becomes wider as $h/R$ 
decreases. In the outer $q$ range,  as expected, the surface approximation no 
longer works since the intensity behaves as $1/q^4$ in agreement with Porod's 
law.  Fig.\,\ref{Fig5}b allows one to better appreciate how the agreement 
between the particle and the surface-like intensities improves throughout 
$0<q<2\pi/h$ as $h/R$ 
decreases.  
%%%%%%      INETNSITY BEHAVIOUR IN THE THREAD-LIKE CASE   %%%%%%
%%%%%%     Intensity behavior in the  thread-like case       %%%%%%%%%%
\subsection{Intensity behavior in the  thread-like case}
Though, in most cases, a thread-like phase exists in presence of  another single 
phase, hereafter  one assumes that it exists in presence of two further phase and 
that the thread-like phase portion that lies on the interface between  
phases 2 and 3 is negligible so as to have $L=L_2+L_3$, where $L_2$ and $L_3$ 
respectively denote the total lengths of the threads that fully lie within phases 2 
and 3. It is also assumed that 
\begeq\label{5.7.a}
S_{2,3}>>S_{1,2}\quad{\rm and}\quad S_{2,3}>>S_{1,3}
\endeq and that the mean of the curvature radii of $S_{2,3}$ is not smaller than 
$\bRc$, defined by (\ref{4.7}). The above assumptions amount to say that the 
threads are in an extended configuration and that the mean distance among 
them is not smaller than $\bRc$. These properties make it possible to apply 
the considerations made  in \S\,5.1 to the thread-like case. Thus,  the equation 
equivalent  of  (\ref{5.3d}) and (\ref{5.3b}b)  becomes 
\begeq\label{5.3h}
{\tilde{\bar \gamma}}_{\cC}(q) = \frac{4\pi}{q}
\frac{\sigma^2 L}{\phi_1(1-\phi_1)\,V}\int_{2\delta}r\sin(qr)
\gamma_{\cC}(r)\rd r. 
\endeq
By (\ref{4.19}), considering the only leading term, one finds 
\begeq\label{5.3ha1}
{\tilde{\bar \gamma}}_{\cC}(q) \approx 
\frac{2\sigma^2 L}{V\phi_1(1-\phi_1)\,q}\int_{2q\delta}\frac{\sin(x)}{x}
\rd x \approx \frac{2\sigma^2 L}{V\phi_1(1-\phi_1)\,q}
\Big[\frac{\pi}{2}-{\rm Si}(2q\delta)\Big], 
\endeq
where ${\rm Si(x)}$ is the sine integral function  (Gradshteyn \& Ryzhik, 1980) 
that, at small $x$, behaves as $-x$.  
%%                           Using the property that $\phi_1\approx\sigma\,L$, 
One concludes that 
\begeq\label{5.3ha}
{\tilde{\bar \gamma}}_{\cC}(q) \approx{\tilde\Gamma}_1(q)\approx  
\frac{\pi\sigma^2L}{V\,\phi_1 (1-\phi_1)\,q}\quad 
{\rm if}\quad \frac{2\pi}{\bRc}<q<\frac{ \pi}{\delta}.
\endeq 
In this $q$-range the $\tilde\Gamma_i(q)$s with $i=2,3$ are negligible since 
they are $O(\delta^3)$. In the range $q>2\pi/\delta$ the behaviors are similar 
to those found in the film-like case. One finds 
\begeq\label{5.3hb}
  {\tilde\Gamma}_1(q)\approx  \frac{2\pi\,(S_{1,2}+S_{1,3})}{V\phi_1(1-\phi_1)\,q^4},
\quad\quad\quad\quad\quad\quad   {\rm if}\quad \frac{ \pi}{\delta}<q, 
\endeq 
and 
\begeq\label{5.3hc}
  {\tilde\Gamma}_i(q)\approx  \frac{2\pi(S_{1,i}+S_{1,j})}{V\phi_i(1-\phi_i)\,q^4},
\quad i\ne j=2,3\ \quad  
{\rm if}\quad \frac{2\pi}{\bRc}<q.
\endeq
By (\ref{5.3ha}),   (\ref{5.3hb}),   (\ref{5.3hc}) and  (\ref{5.1}) the 
asymptotic behavior of  the thread-like scattering intensity is  
\begeq\label{5.1.8a}
I(q)\approx 
\begin{cases}\frac{\cP_{\cC}}{q}\ \  \  &\text{ if } \quad 
\frac{2\pi}{\bRc}<q<\frac{\pi}{\delta},  \\
 \frac{\cP}{q^4}\ &\text{ if } \quad \frac{\pi}{\delta}<q.
\end{cases}
\endeq   
with 
\begeq\label{5.1.8ab}
\cP_{\cC}\equiv \frac{\pi\, n(1;2,3)\sigma^2\,L}{2}= 
\frac{\pi\, n(1;2,3)\sigma\,\phi_1 V}{2}
\endeq 
and $\cP$ given by (\ref{5.1.5abx}). \\ 
\begin{figure}[!h]                   %%       FIGURE                    %%%%%%%%%%%%%    
{\includegraphics[width=7.truecm]{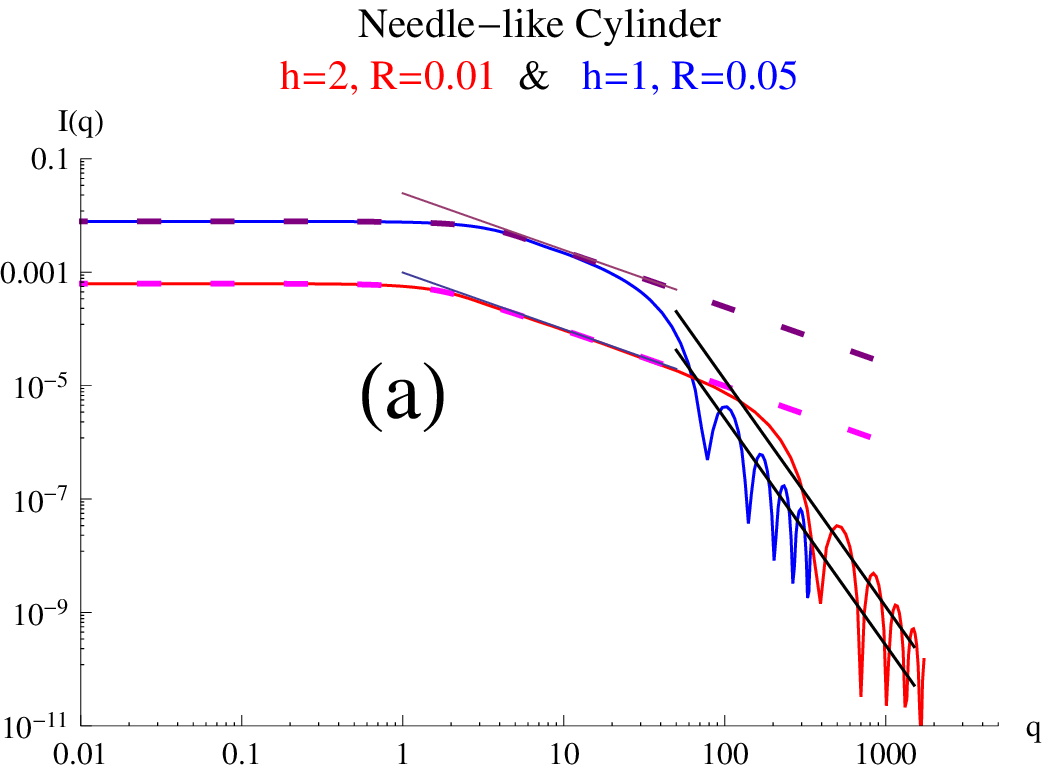}}
{\includegraphics[width=7.truecm]{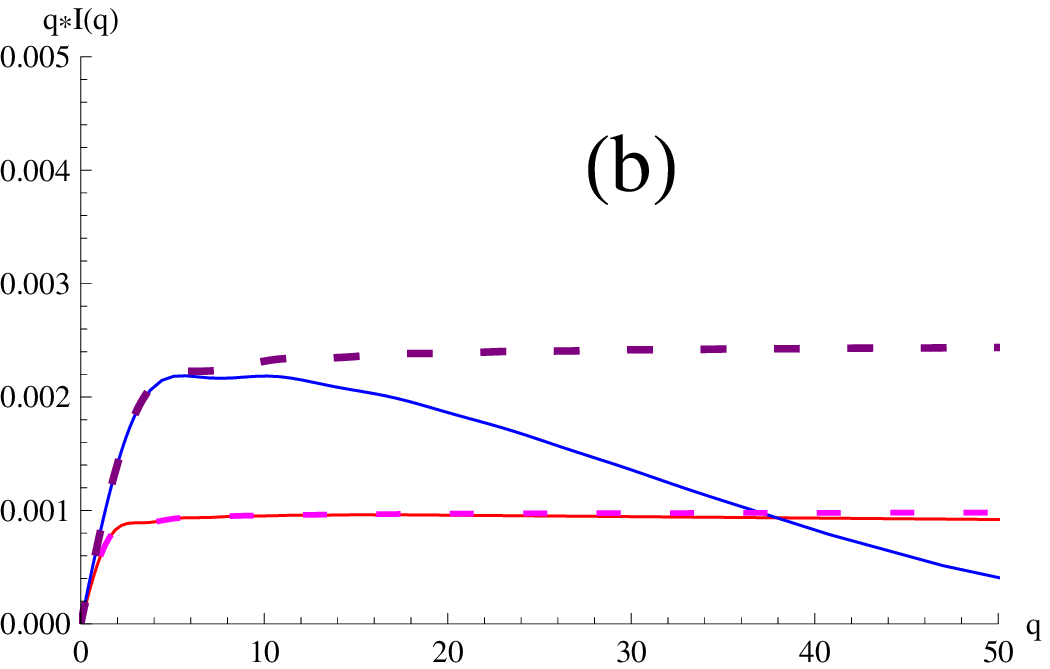}}
\caption{\label{Fig6} { The two panels, similarly to Fig. \ref{Fig5}, show  the 
log-log plot and the $qI(q)$ versus  $q$ plots of the intensities relevant to two 
needle-like cylinders of  height ($h$) and radius ($R$)  as specified at the
top of panel (a) as well as their thread-like approximation given by (\ref{5.1.9}). }}
%%%%   SEE THE FILE  %% ""/Users/salvino/Desktop/WORK_IN_PRGS/THREAD_FILM_LIKE_SYSTEMS"/FTs_cylinder"
\end{figure}
The main consequence of this analysis is that the {\em log-log} plot of 
the scattering intensity of a thread-like sample may show a linear behavior 
with slope -1 at intermediate $q$ values and another linear behavior with 
slope -4 (\ie\,the Porod one) at larger $q$s. The associated constants 
$\cP_{\cC}$ and $\cP$ are related to the the section and the length of 
the thread-like phase, the scattering contrasts and the interface surface 
areas as reported in (\ref{5.1.8ab}) and (\ref{5.1.5ax}).  The expressions 
immediately convert to the two phase case ones  by setting $n_3=n_2$ 
and $S_{2,3}=0$. In this way the  $\cP_{\cC}$ expression reduces to that 
obtained by Kirste \& Oberth\"ur (1988)  under the more restrictive 
assumption that the particles are rod-like.  In particular, in the two phase 
case, the determination of $\cP_{\cC}$  and $\cP$ allows one to determine 
the normal section and the total surface areas of the thread-like phase if 
one knows the contrast because  volume fraction $\phi_1$ can be obtained 
by the Porod invariant value [\ie\, equation (\ref{5.3})].  \\  
Figure \ref{Fig6} reports an illustration of these results considering the 
case of a needle-like cylinder, \ie\, a cylinder with its height $h$ larger 
than is diameter $2R$.  The scattering intensity is obtained by the numerical 
integration of (\ref{A.6}). The intensity associated to the limit $\sigma\to 0$ 
of the thread like phase is obtained, according to (\ref{4.12}), multiplying 
by $h\sigma^2/V$ the FT of (\ref{4.20}). In this way one finds   
\begeq\label{5.1.9}
\frac{h\sigma^2}{V}\,{\tilde\gamma}_{\cC,ls}(q)=\frac{2\sigma}{h\,q^2}
\Big[\cos(hq)+(hq)\,{\rm Si}(hq)-1\Big],
\endeq
where ${\rm Si}(x)$ is the sine integral function. 
The expression on the right hand side approaches  $\pi\,h\,R^2$ as $q\to 0$ and 
behaves as $\pi^2 R^2/q$ in the range $qh>>1$. The figure shows that the region 
where the intensity behaves as $1/q$ enlarges as the $R/h$ ratio decreases. 
Actually, as in the previously reported cases,  this value must be smaller than  few 
percents for the $q^{-1}$ behavior to be clearly observed.  
%%%%%%                       {The right parallelepiped case}
\subsection{The right parallelepiped case}
%%%%%%                       {The right parallelepiped case}
It may happen that a given sample appears as made up of threads if observed 
on a coarse length scale and of films  if observed on a fine one. Then, 
its scattering intensity will show both a $q^{-1}$ and a $q^{-2}$ behavior, 
besides the $q^{-4}$ one at very large $q$s. A dilute monodisperse and 
statistically isotropic sample made up of right parallelepipeds of sizes 
$a\times b\times c$ with $a<< b<<c$,  is a paramount example of this 
phenomenon, as it will now be shown. The monodispersity assumption 
allows one to confine the attention to the behaviors of the CF and the form 
factor of a single parallelepiped.   The CF of this particle shape  was  worked 
out by Gille (1999) and has an analytic form. Since one expects that  the 
parallelepiped looks as a linear segment  on a length scale greater than $b$ 
(and smaller than $c$) and as a rectangle on a scale greater than $a$ 
(and smaller than $b$), from equations (\ref{4.12}) and (\ref{3.6}) it follows 
that 
\begeq\label{5.3.1}
\gamma_{V,prl}(r)\approx\begin{cases}
\frac{(ab)^2c}{abc}\gamma_{\cC,ls}(r,c)  &\text{if} \quad    b<r<c\\
\frac{a^2\,bc}{abc}\gamma_{\cS,rct}(r;b,c)  &\text{if}\quad    a<r<b,   
\end{cases}
\endeq
where $\gamma_{V,prl}(r)$ is the parallelepiped CF  reported by Gille (1999), 
$\gamma_{\cC,ls}(r,c)$  the linear segment CF given by (\ref{4.20}) and 
$\gamma_{\cS,rct}(r;b,c)$ the rectangle surface CF given by equation (\ref{5.3.2}). 
\begin{figure}[!h]                   %%       FIGURE                    %%%%%%%%%%%%%    
{\centering{\centering{\includegraphics[width=7.truecm]{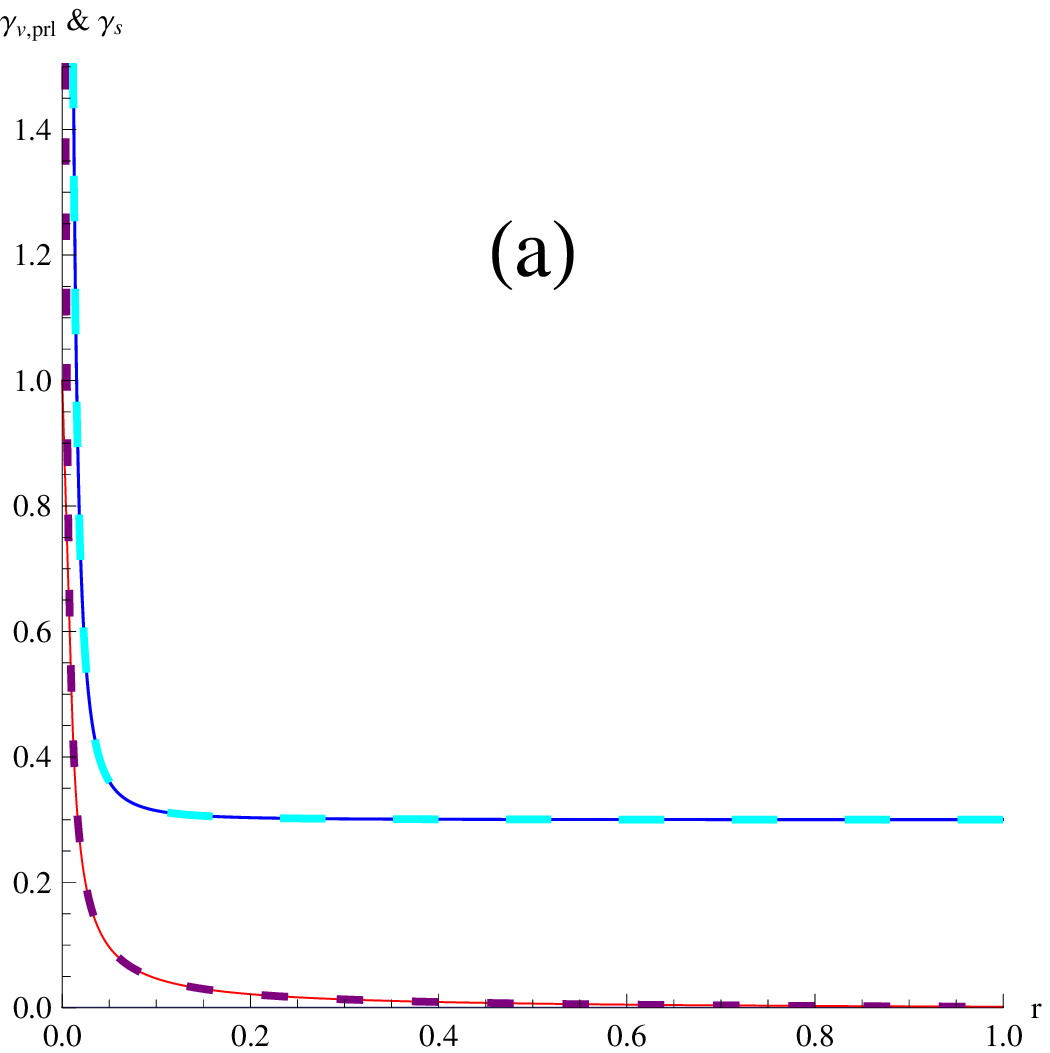}}}} 
%"parallele.eps"%%%%old_name
{\includegraphics[width=7.truecm]{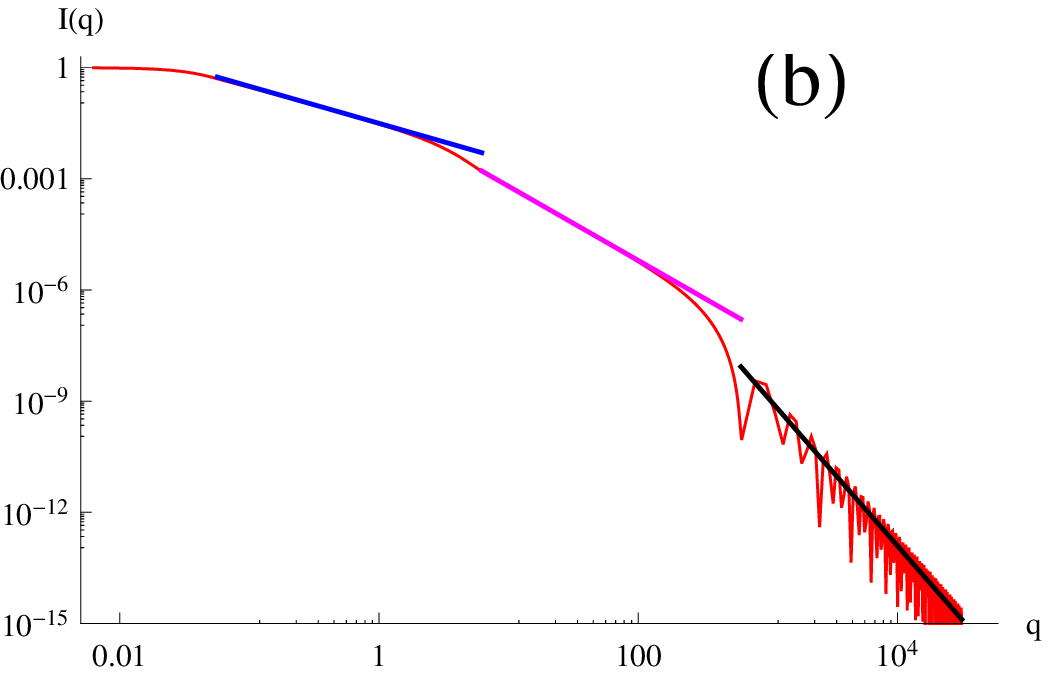}}  %" parallele.eps" old_name
{\includegraphics[width=7.truecm]{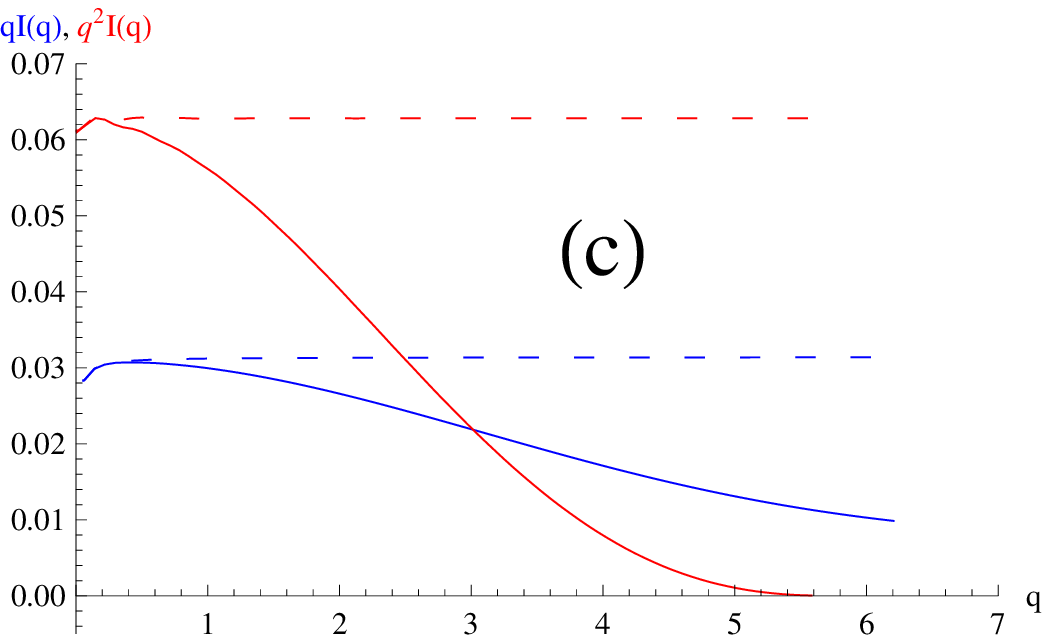}}   %"prllpdPordPlt.eps" old_name
\caption{\label{Fig7} {(a) Behavior of the parallelepiped CF  and its thread  
and film-like approximations; (b)  behavior of the  scattering intensitiy and its 
leading   $q^{-1}$, $q^{-2}$ and $q^{-4}$  "asymptotic" terms; (c) behavior of 
the $qI(q)$ (blue)  and $q^2I(q)$ (red) versus $q$ plots of the parallepiped 
intensity (continuous curves) and of the relevant curve and surface intensity 
approximations (broken curves).  }}
\end{figure}
Figure \ref{Fig8}a shows the accuracy achieved by  (\ref{5.3.1}) in approximating 
the CF of a right parallelepiped with $a=1/100$u, $b=1$u and $c=100$u by 
the relevant film and thread CFs. In fact, the continuos red curve is the plot of 
$\gamma_{V,prl}(r)$ and the broken magenta one that of 
$a\cdot\gamma_{\cS,rct}(r;b,c)$ throughout the range $a<r<b$. The 
continuous blue curve is the plot of $10^3\cdot \gamma_{V,prl}(r)$ 
and the broken cyan one that of $10^3\cdot ab\cdot\gamma_{\cC,ls}(r,c)$ 
throughout $b<r<c$ (this interval has been linearly mapped over 
[0.01,\,1], and both curves are vertically shifted by 0.3). The form factor 
$I_{V,prl}(q)$ of the considered parallelepiped is the red curve 
shown in  fig. 8b  and is  numerically obtained evaluating the FT of  
$\gamma_{V,prl}(r)$. The blue, the magenta and the  black linear segments,  
which respectively behave as $q^{-1}$,  $q^{-2}$ and $q^{-4}$,  have been 
obtained by equations (\ref{5.1.8a}a),  (\ref{5.1.5}a) and  (\ref{5.1.8a}b)  [or 
(\ref{5.1.5}b)]. They have been drawn within the $q$-ranges:  
$[2\pi/c,2\pi/b]$, $[2\pi/b,2\pi/a]$ and $[2\pi/a, 6\cdot10^4u^{-1}]$.  
One sees that the $q^{-1}$ and $q^{-2}$ behaviors are fairly obeyed close 
to the lower bounds of the previous two $q$ intervals. In the proximity of the 
two upper bounds, the intensity deviates from the reported two linear behaviors 
as it appears more evident in figure \ref{Fig8}c. Here the continuous and broken
 blue are the plots of $q\,I_{V,prl}(q)$  and of   $abq{\tilde \gamma_{C,ls}}(q,c)$ 
[note that, similarly to the previously reported cases,  this FT  has been  
evaluated integrating over the full support of $\gamma_{C,ls}(r,c)$ to 
ensure its positiveness] within the $[2\pi/c,2\pi/b]$ $q$-range, 
and the continuous and broken red  curves are the Kratky plots of $ I_{V,prl}(q)$ 
and the FT of $ab{\tilde \gamma_{\cS,rct}}(q;b,c)$ (evaluated over 
$[0,\sqrt{b^2+c^2}])$ within $[2\pi/b<q<2\pi/a]$ (the last interval 
has been linearly mapped over the former one). 
%%%%%%%%%%%%%%     %%%%%%    CONCLUSION   %%%%%%%%%%%%%%%%%%
\section{Conclusion} 
The reported analysis has shown that the CF of a thread-like or a film-like 
statistically isotropic  sample  can be approximated by a 3rd degree 
polynomial in an inner distance range %$0<r<2\delta$ [or $\delta$] 
and, externally to this,  by the associated curve or surface CF.
% in the outer distance range. 
The curve 
and the surface CFs respectively behave as $1/r^2$ and as $1/r$ close to 
the origin. Consequently,  the relevant scattering intensities respectively 
behave as $\cP_{\cC}/q$ or as $\cP_{\cS}/q^2$ in a range of intermediate $q$s 
and as $\cP/q^4$ in the outer $q$ range. The $\cP$ and $\cP_{\cS}$ expressions 
were determined by Porod (1951) and by Teubner (1990). The expression of  
$\cP_{\cC}$ is new. Both 
$\cP_{\cS}$ and $\cP_{\cC}$ coincide with the expressions obtained by 
Porod (1982) and by Kirste \& Oberth\"ur (1982) in the cases of plane lamellae and 
circular rods.  On a practical ground, coefficients $\cP$ and $\cP_{\cC}$ 
 or  $\cP_{\cS}$ can easily be determined from the linear portions of 
the intensity log-log plot. Their knowledge determines, in a (nearly) model
 independent way, the interface area as well as  the normal thread section 
area or the film thickness. Furthermore, a numerical check on the physical 
consistency of the assumed film-like or thread-like structure is possible because 
the lower bound of the $q$-range where the $q^{-2}$ or $q^{-1}$ behavior occurs 
must be close to the film thickness value or  to the thread maximal chord value 
(estimated from the normal section area),  respectively determined by  $\cP_{\cS}$ and 
$\cP_{\cC}$. The model illustrations, reported in \S\,3.4.1, 3.4.2 and 
4.1.1, suggest that the linear behaviors in the log-log plots are observable if  
the two typical lengths, $\delta$ and ${\bar R}_s$ for the film-like case and 
$2\delta$ and ${\bar R}_c$ for the thread-like one, differ at least by an order 
of magnitude. This fact is confirmed by Figures 5 and 6. The paper analysis also 
applies to three phase systems. 
In thie case of film-like samples involving three phases, the knowledge  of 
coefficients $\cP_{\cS}$, $\cP$ and $\cQ_P$ only puts some bounds on 
the involved structural parameters since it is not sufficient to uniquely 
determine them.  Whenever the scattering intensity 
is collected over a wider $q$-range, as it happens using also ultra-small scattering 
equipments, one might study samples that behave in a thread-like way in the 
innermost $q$-range and in a film-like one in the intermediate $q$-range   
(see the  model illustration reported in \S\,5.3). In this case, a trivial extension of 
the above analysis  makes it possible to determine both the thread section area 
and the film thickness  of the  analyzed sample. 

%%%%%%%%%%%%%%%%%%%%%%%%%%%%%%%%%%%%%%%%%%%%%%%%%%%%%%%%%%%%
%%%   \section{Appendix A}                             %%%%%%%   
\section*{Appendix A: the  circular cylinder  CF}
%\subsection*{}
%%%   \section{Appendix A}       Circular CYLINDER CF                      %%%%%%% 
%%%%%%%%%%%%%%%%%%%%%%%%%%%%%%%%%%%%%%%%%%%%%%%%%%%%%%%%%%%%
The chord length distribution  (CLD) of a circular right cylinder of radius $R$ and 
height $h$ is since long known (Gille, 2014). It  involves the elliptic integral 
functions $E(\varphi,k)$ and $F(\varphi,k)$ as well as the complete elliptic integrals 
$\bE(k)$ and $\bK(k)$ [we are adopting here Gradsteyn 
\& Ryzhik's (1980) definitions]. Besides, it is known that the CLD takes two different 
form depending on whether the cylinder has a disk-like form (\ie\ $h<2R$) or a 
needle-like one (\ie\ $2R<h$). In the following two subsection one reports the explicit 
expressions of the CF that so far were never written down. \\
To this aim,  one first puts
\begeq\label{A.1}
\Delta_1\equiv \sqrt{4\,R^2-r^2},\quad \Delta_2\equiv \sqrt{r^2-h^2},\quad\Delta_3\equiv \sqrt{4R^2+h^2-r^2}
\endeq
\begeq\label{A.2}
\cG\equiv\frac{1}{24\pi\,r h R^2},\quad\xi\equiv \frac{r}{2\,R},\quad \zeta\equiv \frac{2\, h\, R}{r\, \Delta_3},
%\quad\eta\equiv \frac{\Delta_1}{2\,R},
\endeq
\begeq\label{A.3}
\varphi_1=\arcsin{\,\xi},\quad\varphi_2\equiv\arcsin\,\zeta,
%\quad\varphi_2\equiv\arcsin\,\eta,\quad\varphi_3\equiv\arcsin\,\zeta,
\endeq
\begeq\label{A.4}
\varphi_3\equiv\arcsin{\frac{\Delta_2}{2\,R}},\quad
\varphi_4\equiv\arcsin\frac{r\Delta_3}{2\,h\,R},
%\varphi_4\equiv\arcsin{\frac{\Delta_2}{2\,R}},\quad
%\varphi_5\equiv\arcsin\frac{r\Delta_3}{2\,h\,R},\quad V_c\equiv \pi\,R^2\,h,
\endeq
\begin{eqnarray}\label{A.4a}
&&G_{A}(r,R,h)\equiv\cG\Big[3\Big(r\big(4\pi R^2(2h-r)+(r^2+2R^2)\Delta_1)+ 
 \\
&&\quad 8R^2(r^2-R^2)\varphi_1\Big)-16h\,R\big((r^2+4R^2)\bE(\xi)+(r^2-4R^2)\bK(\xi)\big)
\Big]  \nonumber
\end{eqnarray}
and 
\begin{eqnarray}\label{A.4b}
&&G_{C}(r,R,h)\equiv \cG\Big[12 \pi R^2 (h^2 + r^2 - R^2) -  \\ 
&&  \quad\quad
(h^2 - 5 r^2 - 26 R^2) \Delta_2\Delta_3-    24 R^2 (h^2 + r^2 - R^2)\varphi_3 -
\nonumber   \\
&&\quad\quad  8 h r\Big((r^2 + 4 R^2)\bE(\varphi_4,\frac{1}{\xi})-
(r^2 - 4 R^2)\bF(\varphi_4,\frac{1}{\xi})\Big)\Big].\nonumber
\end{eqnarray}
%%%%   \end{document}
\subsection*{The disk case}
Then, integrating twice the CLD expression (see the deposited part), in the disk case  
[\ie\ $h<2R$] one finds that the cylinder CF  reads 
\begeq\label{A.5}
\gamma_{dsk}(r,R,h)=
\begin{cases}G_{A}(r,R,h) & \text{if }  0 < r<h,\\
\cG\Big[12\pi h^2 R^2+3r(r^2+2R^2)\Delta_1 +& \\
\frac{\Delta_2}{\Delta_3}(3r^4 - h^4+6 r^2(h^2-R^2)+22h^2 R^2-24R^4)+& \\
24R^2\bigl((r^2-R^2)\varphi_1 -(h^2 + r^2 - R^2)\varphi_3\bigr)- &\\
16 h R \big((r^2 + 4 R^2)\bE(\varphi_2,\xi) + 
 (r^2 - 4 R^2)\bF(\varphi_2,\xi)\big)\Big]&  \text{if }  h < r<2\,R,\\
G_{C}(r,R,h)&  \text{if } 2R < r<\sqrt{4R^2+h^2},\\
0&  \text{if } \sqrt{4R^2+h^2} < r.
\end{cases}
\endeq
\subsection*{The needle case}
Similarly, in the needle case  [\ie\ $2R<h$],  the CF   is found to be 
\begeq\label{A.6}
\gamma_{ndl}(r,R,h)=
\begin{cases}G_{A}(r,R,h) & \text{if }  0 < r<2R,\\
4\cG\Big[3\pi\,R^2(2\,h\, r-R^2)-&\\
2\,h\,r\Big((r^2+4R^2)\bE(\frac{1}{\xi})-
(r^2-4R^2)\bK(\frac{1}{\xi})\Big)\Big]& \text{if }  2R < r<h,\\
G_{C}(r,R,h) &  \text{if } h < r<\sqrt{4R^2+h^2},\\
0 &  \text{if }\  \sqrt{4R^2+h^2}< r.
\end{cases}
\endeq
%%%%                                         \end{document}
%%%%%%%%%%%%%%%%%%%%%%%%%%%%%%%%%%%%%%%%%%%%%%%%%%%%%%%%%%%%
%%%   The surface CF of a cubic surface                             %%%%%%%   
\section*{Appendix B: the cubic surface CF} 
%\subsection*{The surface CF of a cubic surface}  
In deriving equation (\ref{3.11}) it was implicitly assumed that the given surface  
intersects its translated image along a curve. In the reality it can happen that 
for some translation directions the intersection be a surface subset. An illustration 
of this phenomenon is shown in Figure \ref{Fig8} that refers to a cubic 
surface of area $S_f=6 a^2$.  The red polygon, shown in Fig. 8a, is the intersection set of the 
outset cubic surface with its image resulting by the translation of the cubic surface by $r\hw$. 
This vector is such that its tip point does not lie over one of the cube faces.  
The case where $r\hw$ fully lies over one of the cube faces leads to an intersection set 
that is formed by two surfaces that respectively have their borders equal to the red thick 
continuous and  broken rectangles shown in Fig.8b. When theonly  tip of $r\hw$ spans one of the 
cube faces the intersection set is again a surface whose boundary is given by the 
red rectangle shown in Fig. 8c.  To get the CF of the cubic surface it is necessary to 
evaluate integral (\ref{3.7}) imposing the aforesaid constraints on $r\hw$. This task was   
explicitly carried out as detailed in the deposited part of this paper. Here one simply reports 
the surface CF expressions relevant to the geometrical  configurations  illustrated in first three 
panels of Fig. 8. \hfill\break\noindent   
\begin{figure}[hp]
{\includegraphics[width=7.truecm]{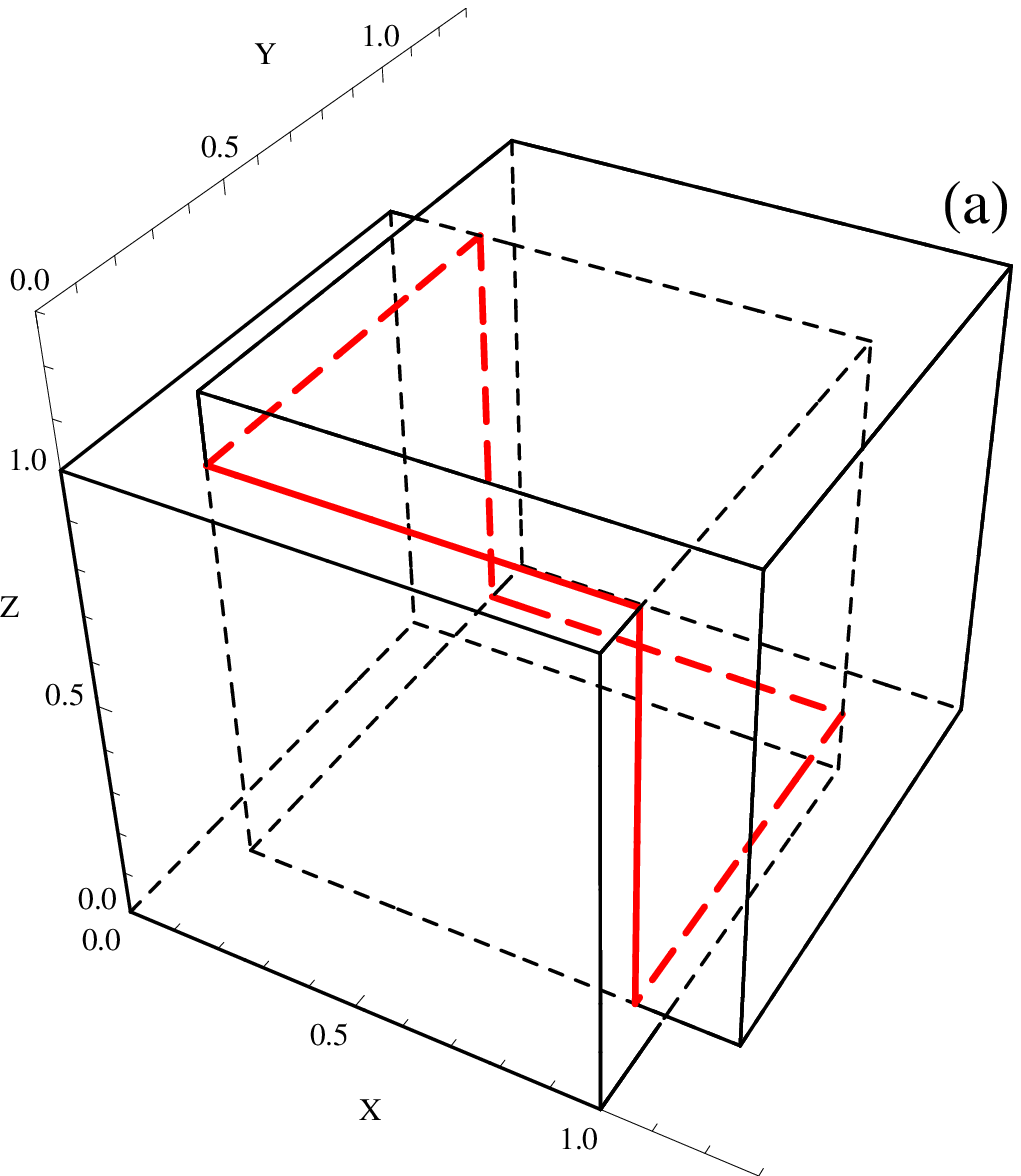}}
{\includegraphics[width=7.truecm]{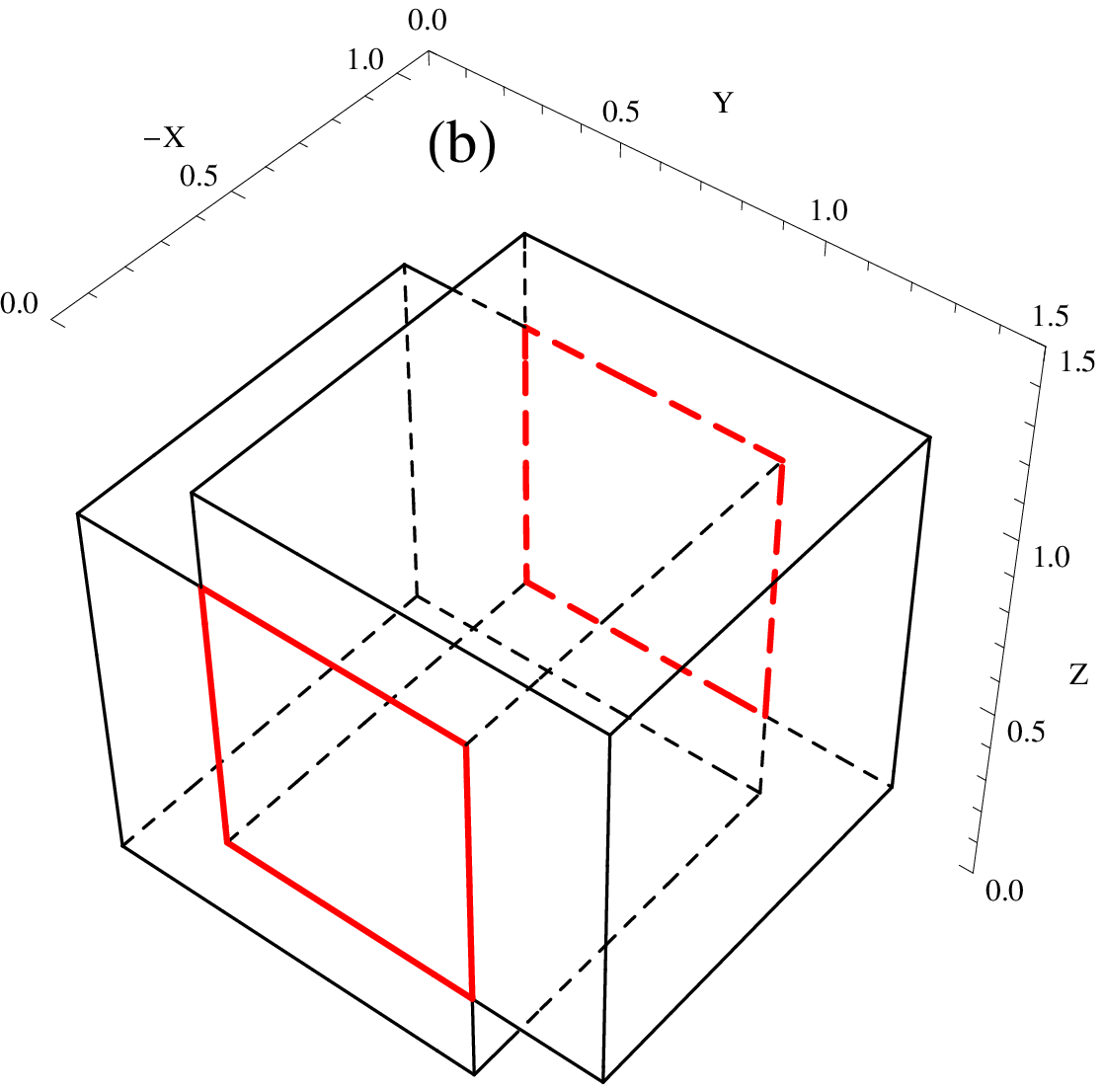}} 
{\includegraphics[width=7.truecm]{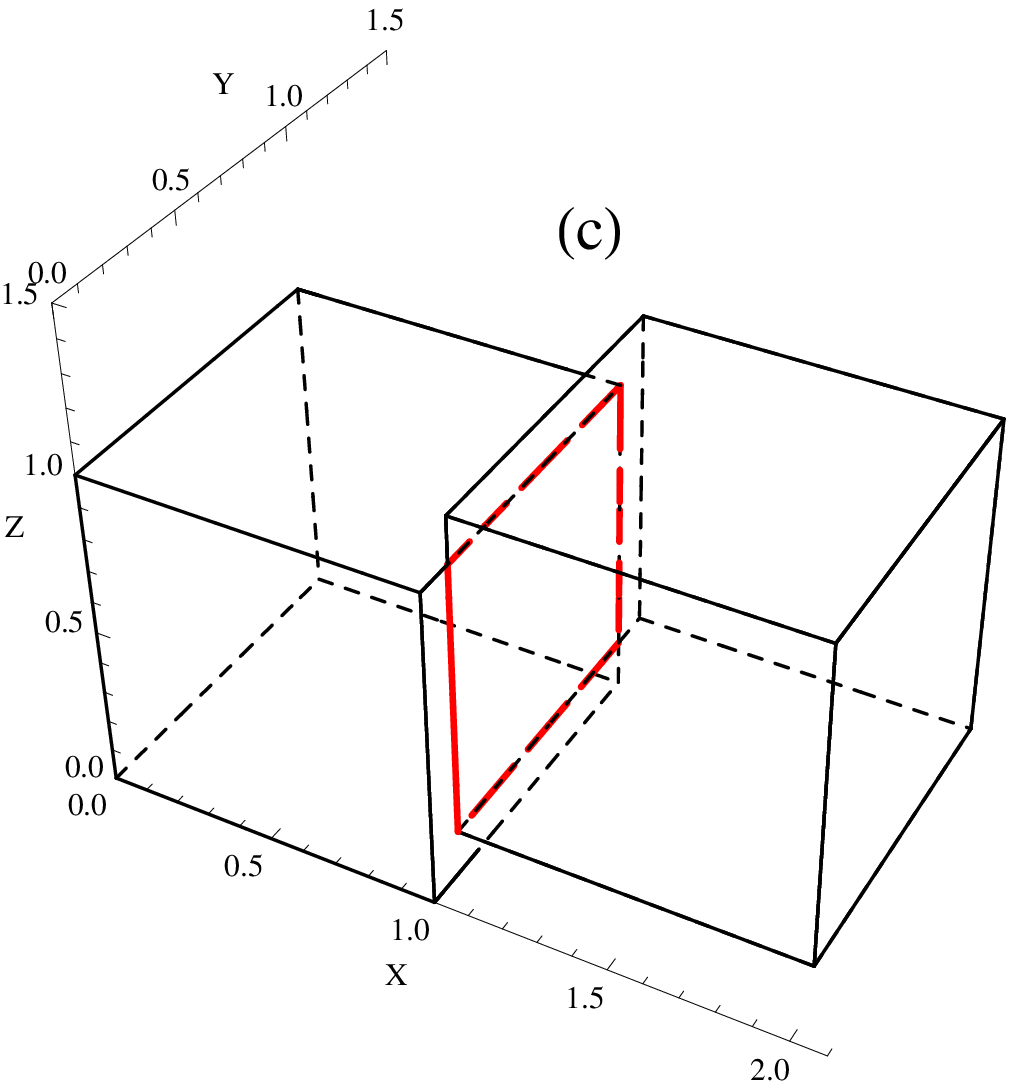}}
{\includegraphics[width=7.truecm]{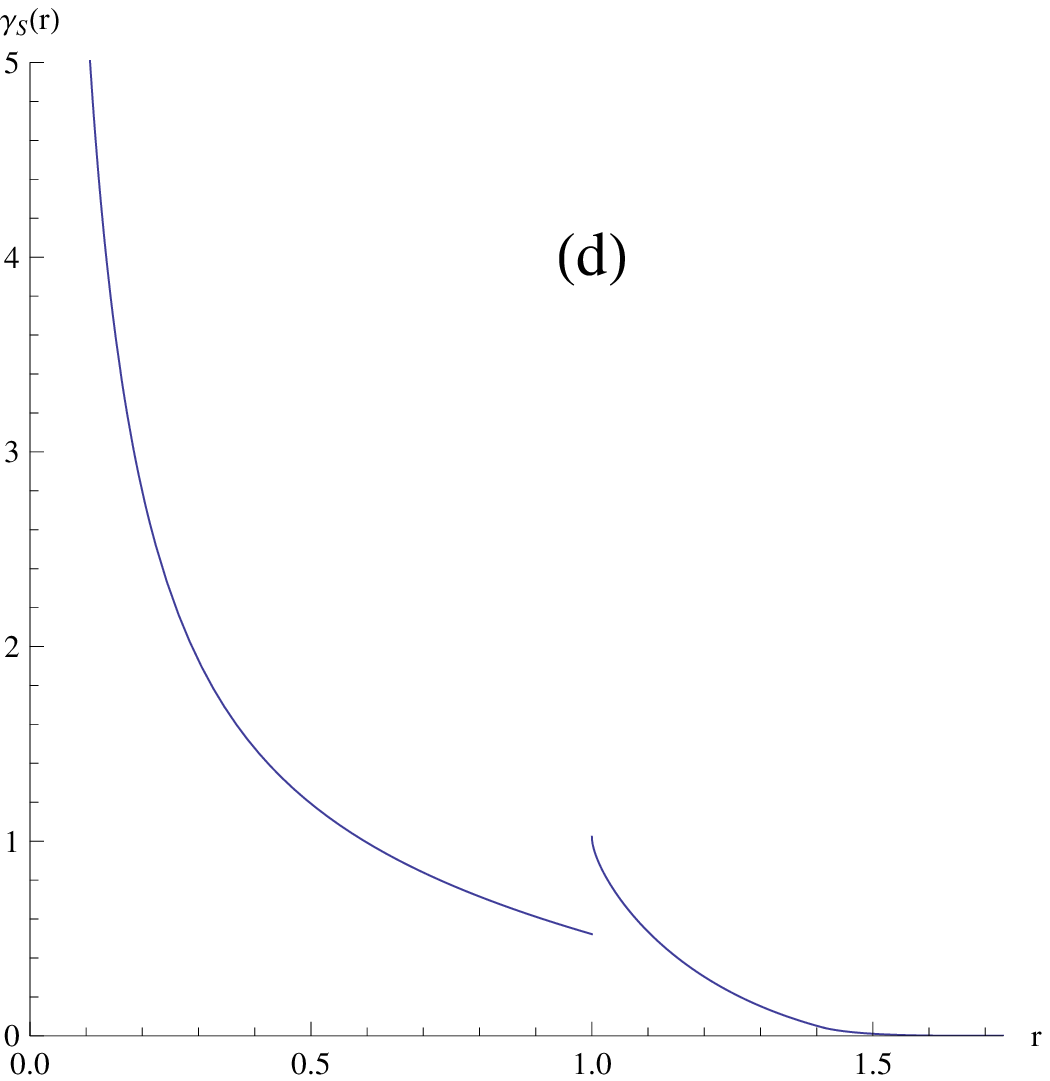}} 
\caption{\label{Fig8} {If $r\hw$ is such that the tip of $r\hw$ does not lie on one of  the 
cube's faces, the intersection of the outset and the translated cubic 
surfaces is made up of the  continuous and broken thick segments shown in panel (a). 
If  $r\hw$ is such that $\theta$ is equal to 0 or $\frac{\pi}{2}$ or $\pi$, so as to have 
two pairs of faces sliding one over the other, the intersection set is formed by two 
opposite rectangles having as borders the continuous and the broken thick red curves 
shown in the (b) panel that explicitly refers to the case $\theta=0$. Finally, if $r\hw$ is such 
that its tip point spans one of the cube's  face, the intersection set is formed by the 
rectangle shared by the superposing faces, \ie\, the rectangle 
having as border the rectangular curve show in red in panel (c). 
Panel (d )shows the final cubic surface CF given by the sum of the aforesaid three 
contributions.  }}
\end{figure} 
They  respectively read 
\begeq\label{3.25X}
\gamma_{\cS,\,CS,\,a}(r,a)\equiv
\begin{cases}
\frac{1}{a} - \frac{r}{2\, a^2}  &\text{if}\ \   0 < r < a,\\  
  \frac{5}{2\, r}-\frac{2}{a}  - \frac{2 \sqrt{r^2-a^2 }}{\pi\,a\, r}+  \\
\frac{2\, r}{\pi\,a^2} \arccos\frac{a}{r} &  \text{ if} \ \    a < r < \sqrt{2}\,a,\\  
\frac{5\pi-12}{6\,\pi\, r}-\frac{2}{a} +\frac{r}{6\,a^2}+
 \frac{2 \sqrt{ r^2-2\, a^2}}{\pi\, a\,r}  +  \\
\frac{4} {\pi\,a}\arcsin\Bigl(\frac{r^2 + a^2}{r^2-a^2}
\sqrt{\frac{r^2-2\, a^2}{2\, r^2-2\, a^2}}
 \Bigr)+\\
\frac{2 (r^2+5 a^2)}{3\,\pi\, a^2\,r}
\arcsin\Bigl( \frac{7\, a^3 - 3\, a\, r^2}{( r^2-a^2 )^{3/2}}   \Bigr)
& \text{if}\  \sqrt{2}\,a<r<\sqrt{3}\,a,\\
0 &\text{if}\quad \sqrt{3}\,a<r,
\end{cases}
\endeq  
\begeq\label{3.25Y}
\gamma_{\cS,\,CS,\,b}(r,a)\equiv 
\begin{cases}
 \frac{1}{2\, r} -\frac{2}{\pi\,a} + \frac{r}{2\pi\,a^2}  &\text{if\ \ }   0 < r < a,\\
-\frac{2+\pi}{2\,\pi\, r}  -\frac{ r}{2\pi\, a^2} + \frac{2 \sqrt{r^2-a^2}}{\pi\,a\,r}+\\
\frac{2}{\pi\,r}\,\arcsin(\frac{a}{r})  &\text{if\ \ }      a < r < \sqrt{2}\,a, \\
0&\text{if}\ \  \sqrt{2}\,a<r.   
\end{cases}
\endeq 
%\end{document}
and 
\begeq\label{3.25Z}
\gamma_{\cS,\,CS,\,c}(r,a)\equiv 
\begin{cases}
0&\text{if\ \ }   0 < r < a,\\
 \frac{\pi-1}{2\,\pi\, r} +\frac{r}{2\,\pi\,a^2} - \frac{2\sqrt{r^2-a^2}}{\pi\,a\,r}  
&\text{if\ \ }   a < r < \sqrt{2}\,a,\\    %%%% 
-\frac{1}{2\,\pi\,  r} -\frac{ r}{2\,\pi\, a^2} + \frac{2}{\pi\,a\,r}
\sqrt{  r^2-2 a^2} -\\ 
\frac{1}{\pi\, r} \,\arcsin\Bigl(\sqrt{\frac{r^2-2\,a^2}{r^2-a^2}}\Bigr)+\\
\frac{1}{\pi\,r} \, \arcsin\Bigl(\sqrt{\frac{a}{r^2-a^2}}\Bigr)  
&\text{if\ \ }     \sqrt{2}\, a < r < a\sqrt{3}\,a, \\
0&\text{if}\ \  \sqrt{3}\,a<r.   
\end{cases}
\endeq 
Thus, the CF function of a cubic surface is the sum of equations  (\ref{3.25X}) , 
 (\ref{3.25Y}) and   (\ref{3.25Z}). Its first moment  is equal to $6a^2/4\pi$, as 
required by equation (\ref{3.26a}) while its plot is shown in panel (d) of Fig. 8 for the case $a=1$. 
The discontinuity, present at $r=a$, arises from the opposite faces that are parallel at distance $a$. 
%%%%%%%%%%%%%%%%%%%%%%%%%%%%%%%%%%%%%%%%%%%%%%%%%%%%%%%%%%%% 
%%%%%%%%%%%%%%%%%%%%%%%%%%%%%%%%%%%%%%%%%%%%%%%%%%%%%%%%%%%%
%%%   \section{Appendix C}                             %%%%%%%   
\section*{Appendix C:  small distance behavior of the curve CF}
%%%   \section{Appendix C}                             %%%%%%% 
%%%%%%%%%%%%%%%%%%%%%%%%%%%%%%%%%%%%%%%%%%%%%%%%%%%%%%%%%%%%   
In order to prove equation (\ref{4.19}) one proceeds by expanding the parametric equation 
of the curve around $\ell=0$ which is the curvilinear coordinate of the point taken as the 
origin of the Cartesian frame. The expansion up to terms $o(\ell^3)$ reads 
\begeq\label{C.20}
\bR(\ell)\approx \Bigl\{ \ell- \frac{\ell^3}{6\, {R_c}^2},\, \frac{\ell^3}{6\, R_c\, R_t},\, 
\frac{\ell^2}{2\, R_c}\Bigr\}.
\endeq
One  straightforwardly verifies that vectors ${\hat\tau}$, ${\hbn}$ and ${\hbb}$, 
obtained by applying definitions (\ref{4.2})  to (\ref{C.20}) 
are mutually orthogonal  unit vectors up to terms $o(\ell^3)$. 
The condition $\bR(\ell)\cdot\bR(\ell)=r^2$ determines the curvilinear abscissa of the 
points that are at distance $r$ from the origin.  As $r\to 0$, the equation is easily solved 
by iteration, putting $l\approx r$ as first step. The solutions are 
\begeq\label{C.21}
\ell_{\pm}\approx \pm r\Big(1+\frac{r^2}{24\,R_c}\Bigr)+O(r^5). 
\endeq
Consider the positive solution. One finds that $\hw(\ell_+)=\bR(\ell_+)/r$ and   
${\hat\tau}(\ell_+)=\frac{d\bR(\ell)}{d\,\ell}\big|_{\ell=\ell_+}$. In this way, by  
equation (\ref{4.17}), one obtains
\begeq\label{C.22}
\cos\theta_+=\hw(\ell_+)\cdot {\hat\tau}(\ell_+)\approx 1 - \frac{r^2}{8 {R_c}^2} +o(r^2).
\endeq
The result for the negative solution is the same. In this way 
result  (\ref{4.19}) is immediately obtained  by (\ref{4.18}).
%%%%%          
\section*{Acknowledgments}
We  gratefully thank Dr. Wilfried Gille for his critical reading of the ms  and for the suggestion 
of investigating the moments of the curve and the surface CFs.
\vfill\eject
\section*{References}
%\leftline {\bf References}          %%%    REFERENCES
\begin{description}
%\item[\refup{}]Abramowitz, M. \& Stegun, I.A. (1970). {\em Handbook of Mathematical Functions}, New York: %Dover. 
\item[\refup{}] Avdeev, M. V., Aksenov, V. L., Gazov\'a, Z., Almásy,  L.,  Petrenko, V. I., 
Gojzewski, H.,  Feoktystov, A. V.,  Siposova, K.,  Antosova, A.,  Timko M. \&  Kopcansky, P. 
(2013). {\em J. Appl. Cryst.} {\bf 46}, 224-233. 
\item[\refup{}]Ciccariello, S.  (1984).  {\em J. Appl. Phys.}  {\bf 56}, 162-167.
\item[\refup{}]Ciccariello, S.  (1989).  {\em Acta Cryst. A}  {\bf 45}, 86-99.
\item[\refup{}]Ciccariello, S.  (1991). {\em Phys. Rev.} A{\bf 44}, 2975-2983.
\item[\refup{}]Ciccariello, S. (1995). {\em J. Math. Phys.} {\bf 36}, 219-246.
\item[\refup{}]Ciccariello, S. (2009). {\em J. Math. Phys.} {\bf 50}, 103527/20.
\item[\refup{}]Ciccariello, S. (2010). {\em J. Appl. Cryst.} {\bf 43}, 1377-1384.
\item[\refup{}]Ciccariello, S. (2014). {\em J. Appl. Cryst.} {\bf 47}, 1866-1881.
%\item[\refup{}]Ciccariello, S.  (2014). {\em J. Appl. Cryst.} {\bf 47}, 1445-1448 
\item[\refup{}] Ciccariello, S. \& Benedetti, A. (1982). {\em Phys. Rev.} B{\bf 26}, 6384-6389.
\item[\refup{}]Ciccariello, S. \& Riello, P. (2007). {\em J. Appl. Cryst.} {\bf 40}, 282-289.
\item[\refup{}]Ciccariello, S.  \& Sobry,  R. (1995). {\em Acta Cryst. A} {\bf 51}, 60-69.
\item[\refup{}]Ciccariello, S., Cocco, G., Benedetti, A.  \&  Enzo, S. (1981). {\em Phys. Rev.} B{\bf 23}, 6474-6485.
\item[\refup{}] Debye, P., Anderson, H.R.  \&  Brumberger, H. (1957). {\em J. Appl.
Phys.} {\bf 20}, 679-683.
\item[\refup{}] Fedorova,  I.S. \& Emelyanov, V.B. (1977). {\em J. Colloid Interface Sci.} {\bf 59}, 106-112.
\item[\refup{}] Feigin, L.A. \&  Svergun, D.I. (1987). {\em Structure Analysis
by Small-Angle X-Ray and  Neutron Scattering}, New York: Plenum Press.
\item[\refup{}]   Gille, W. (1999). {\em J. Appl. Cryst.} {\bf 32}, 1100-1104.
\item[\refup{}]   Gille, W. (2014). {\em Particle and Particle Systems
characterization }, London: CRC.
%\item[\refup{}]Goodisman, J. (1980). {\em J. Appl. Cryst.} {\bf 13}, 132-34.
\item[\refup{}]Goodisman, J. \& Brumberger, H. (1971). {\em J. Appl. Cryst.} {\bf 4},
347-351.
\item[\refup{}] Glatter, O. (1982). {\em Small-Angle X-Ray Scattering}. Edts Glatter, O. \& 
Kratky, O., London: Academic Press. 
\item[\refup{}]Gradshteyn, I.S. \& Ryzhik, I.M. (1980). {\em Tables of Integrals, Series and Products}, New York: Academic Press.
%\item[\refup{}]Guinier, A. (1946). {\em Compt. Rend.} {\bf 223}, 161-162.
\item[\refup{}]Guinier, A. \& Fournet, G. (1955). {\em Small-Angle Scattering of X-rays.} New York: John Wiley.
\item[\refup{}]Kirste, R.G. \& Porod, G. (1962). {\em Kolloid Z.} {\bf 184}, 1-6.
\item[\refup{}] Kirste, R.G. \& Oberth\"ur, R.C. (1982). {\em Small-Angle X-Ray Scattering}. 
Edt.s Glatter, O. \& Kratky, O., London: Academic Press
\item[\refup{}]Kostorz, G. (1979). {\em Neutron Scattering}, Ed.  Kostorz, G.,   
London: Academic Press,  pp 227-289.
\item[\refup{}] Melnichenko, Y.B. \& Ciccariello, S. (2012). {\em  J. Phys. Chem. C} {\bf 116}, 24661-24671. 
\item[\refup{}]M\'ering, J. \& Tchoubar, D. (1968). {\em J. Appl. Cryst.} {\bf 1}, 153-65.
%\item[\refup{}]Moore, P.B.  (1980). {\em J. Appl. Cryst.} {\bf 13}, 168-175.
%\item[\refup{}]Pedersen, J.S.  (1994). {\em J. Appl. Cryst.} {\bf 27}, 595-608.
\item[\refup{}] Peterlin, A. (1965). {\em Makromol. Chem.} {\bf 87}, 152-160.
\item[\refup{}]Porod, G. (1951). {\em Kolloid Z.} {\bf 124}, 83-114.
\item[\refup{}]Porod, G. (1967). {\em Small-Angle X-Ray Scattering. Proceedings of the 
Syracuse Conference}, Ed.  H. Brumberger, 1-8, New York:  Gordon \& Breach.
\item[\refup{}]Porod, G. (1982). {\em Small-Angle X-Ray Scattering}. Edt.s Glatter, O. \& 
Kratky, O., London: Academic Press. 
\item[\refup{}] Smirnov, V.I. (1970). {\em Cours de Math\'ematiques Sup\'erieures}, 
Moscow: Mir, Vol. II, Chap. V.1.
%\item[\refup{}]Roess, L.C. (1946). {\em J. Chem. Phys.} {\bf 14}, 695-697.
%\item[\refup{}]Roess, L.C. \& Shull, C.G. (1947). {\em J. Appl. Phys.} {\bf 18}, 308-313.
%\item[\refup{}]Taupin, D.  \& Luzzatti, V. (1982). {\em J. Appl. Cryst.} {\bf 15}, 289-300.
\item[\refup{}] Teubner, M. (1990). {\em J. Chem. Phys.} {\bf 92},  4501-4507.
\item[\refup{}] Wu, H. \& Schmidt, P. W. (1974). {\em J. Appl. Cryst.} {\bf 7}, 131-146.
%\item[\refup{}]Vonk, C.G. (1976). {\em J. Appl. Cryst.} {\bf 9}, 433-440. 
\end{description}
\end{document}